\documentclass[12pt,onecolumn, draftclsnofoot]{IEEEtran}
\usepackage{graphicx,subfigure}
\usepackage{amsmath,amssymb}
\usepackage{algorithm, algorithmic}
\usepackage{amsthm}
\usepackage{yhmath}
\usepackage{multirow}

\newtheorem{observation}{Observation}

\begin{document}

\title{On Connectivity of Airborne Networks}
\author{\IEEEauthorblockN{ Shahrzad  Shirazipourazad, Pavel Ghosh and Arunabha Sen}\\
\IEEEauthorblockA{ \small Computer Science and Engineering Program\\
\small School of Computing, Informatics and Decision System Engineering\\
\small Arizona State University\\
\small Email: \{sshiraz1, pavel.ghosh, asen \}@asu.edu}}

\maketitle
\vspace{-.4 in}
\begin{abstract}
Mobility pattern of nodes in a mobile network has significant impact on the connectivity properties of the network. One such mobile network that has drawn attention of researchers in the past few years is the Airborne Networks (AN) due to its importance in civil and military purpose and due to the several complex issues in these domains. 
Since the nodes in an airborne network (AN) are heterogeneous and mobile, the design of a reliable and robust AN is highly complex and challenging. In this paper a persistent backbone based architecture for an AN has been considered where a set of airborne networking platforms (ANPs - aircrafts, UAVs and satellites) form the backbone of the AN. End to end connectivity of the backbone nodes is crucial in providing the communication among the hosts. Since ANPs are prone to failure because of attacks like EMP attack or jamming, another important issue is to improve the robustness of the backbone network against these attacks. Such attacks will impact specific geographic regions at specific times and if an ANP is within the fault region during the time of attack, it will fail. This paper focuses on connectivity and fault-tolerance issues in ANs and studies algorithms to compute the minimum transmission range of ANPs in fault free and faulty scenarios to ensure network connectivity all the times. It also considers the scenarios where the network may have to operate in a disconnected mode for some part of time and data transmissions may be tolerant to some amount of delay. Hence, ANPs may not need to have end-to-end paths all the time but they should be able to transmit data to each other within bounded time. 
\end{abstract}

\section{Introduction}
\label{sec:intro}

\begin{figure}[!t]
\centering
\includegraphics[width=0.6\textwidth, keepaspectratio]{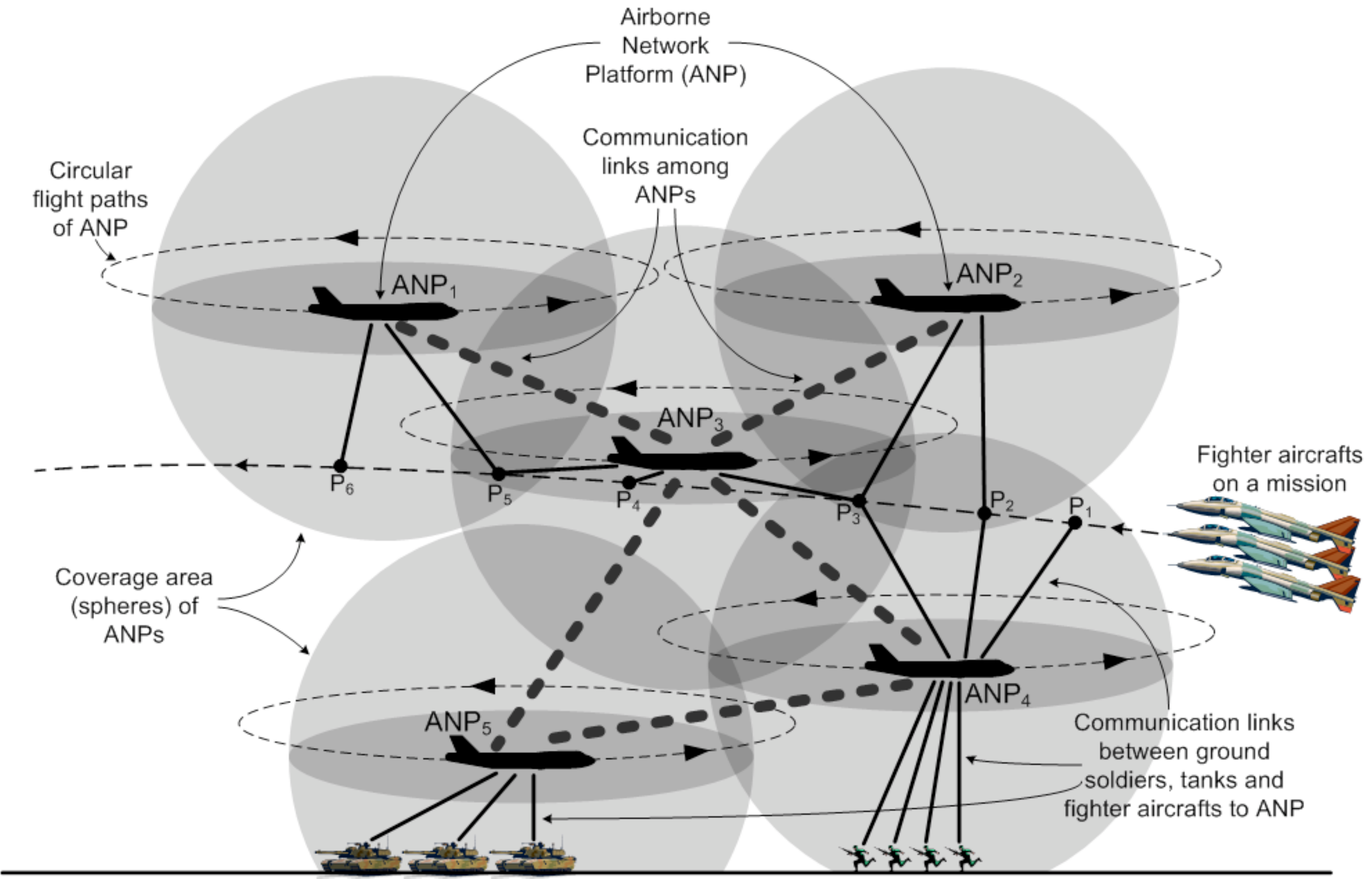}
\caption{A schematic view of an Airborne Network }
\label{fig:airborne}
\end{figure}

An Airborne Network (AN) is a mobile ad hoc network that utilizes a heterogeneous set of physical links (RF, Optical/Laser and SATCOM) to interconnect a set of terrestrial, space and highly mobile airborne platforms (satellites, aircrafts and Unmanned Aerial Vehicles (UAVs)). Airborne networks can benefit many civilian applications such as air-traffic control, border patrol, and search and rescue missions. The design, development, deployment and management of a network where the nodes are mobile are considerably more complex and challenging than a network of static nodes. This is evident by the elusive promise of the Mobile Ad-Hoc Network (MANET) technology where despite intense research activity over the previous years, mature solutions are yet to emerge \cite{Burbank06, Con07}. One major challenge in the MANET environment is the unpredictable movement pattern of the mobile nodes and its impact on the network structure. In case of an AN, there exists considerable control over the movement pattern of the mobile platforms. A senior Air Force official can specify the controlling parameters, such as the {\em location, flight path and speed} of the ANPs, to realize an AN with desired functionalities. Such control provides the designer with an opportunity to develop a topologically stable network, even when the nodes of the network are highly mobile. 
 It is increasingly being recognized in the networking research community that the level of {\em reliability} needed for continuous operation of an AN may be difficult to achieve through {\em a completely mobile, infrastructure-less network} \cite{Milner03}
. In order to enhance {\em reliability} and {\em scalability} of an AN, Milner {\em et al.} in \cite{Milner03} suggested the formation of a {\em backbone network} with Airborne Networking Platforms (ANPs). 
In order to deal with the reliability and scalability issues in an AN
, we consider an architecture for an AN where a set of ANPs form the {\em backbone} of the AN. 
 This set of ANPs may be viewed as {\em mobile base stations} with {\em predictable and well-structured flight paths} and the combat aircrafts on a mission as {\em mobile clients}. We want that the backbone network remains connected all the times even though the topology of the network changes with the movement of the ANPs. 
 Network connectivity can be easily achieved if the transmission range of the ANPs is very large. However large transmission range also implies high power consumption. In order to minimize power consumption and hence extend network lifetime, we would like to find the smallest transmission range to ensure network connectivity.
We define the {\em critical transmission range} ($CTR$) to be the minimum transmission range of the ANPs to ensure that the dynamic network formed by the movement of the ANPs remains connected at all times. We present an algorithm to compute $CTR$ when the flight paths are known. As a part of design of this algorithm, we develop techniques to compute the {\em dynamic topology} of the AN at any instance of time.

Using $CTR$ as the transmission range of all nodes, the network is connected as long as all the network nodes (i.e., the ANPs) are operational. However, the ANPs are vulnerable to Electromagnetic Pulse (EMP) attacks or jamming. Such an attack will impact specific geographic regions at specific times and if an ANP is within the fault region during the time of attack, it will not be able to carry out its normal communication functions.  We will refer to these ANPs as faulty nodes of the network. In this research, we also consider the AN scenario where some of the network nodes are faulty.  We consider faulty nodes are {\em spatially correlated} (or {\em region-based}), that is faulty nodes due to an attack are confined to a {\em region}. We want that the network remains connected {\em irrespective of location of the fault region} and the {\em time of failure}. 

 We define {\em critical transmission range in faulty scenario ($CTR_f$)} to be the {\em smallest transmission range necessary to ensure network connectivity, irrespective of (a) the location of the fault region and (b) the time of the failure}. We would like to find $CTR_f$. As a part of design of this algorithm, we develop techniques to (i) compute all the fault regions that need to be considered to {\em ensure overall connectivity at all times}, (ii) compute the set of  dynamic nodes that might be affected by the {\em failure of a specific region at a specific time}, and finally, (iii) compute $CTR_f$.  

In previous problems the connectivity requirement is very strict and the backbone network is needed to be connected all the times. However, it may not be possible to equip the ANPs with radios 
with transmission range at least as large as the $CTR$. In such a scenario the network may operate in a disconnected mode for some part of time. On the other hand, based on the type of data that should be transmitted between ANPs, data transmissions may be tolerant to some amount of delay. Hence, ANPs may not need to have end-to-end paths all the time but they should be able to transmit data to each other within bounded time. These requirements lead us to study the problem of computation of critical transmission range in delay tolerant airborne networks. More specifically, the critical transmission range in delay tolerant network ($CTR_D$) is defined to be the minimum transmission range necessary to ensure that every pair of nodes in the backbone network can communicate with each other within a bounded time. In this paper we formulate $CTR_D$  and propose a solution to compute $CTR_D$.

The rest of the paper is organized as follows. We discuss related work in  section \ref{sec:relatedworks}.  In section \ref{sec:sysmodel} we present the AN architecture. We present dynamic topology computation of the AN in section \ref{sec:dynamictopology}. In section \ref{sec:faultfree} we present an algorithm to compute $CTR$ in fault free scenario. We discuss the faulty scenario and propose an algorithm to compute $CTR_f$ in section \ref{sec:faulty}. Connectivity problem in delay tolerant airborne network is formulated and studied in section \ref{sec:DTN}. Experimental results are presented in section \ref{sec:simulation}. The section \ref{sec:conclusion} concludes the paper.

\section{Related Works}
\label{sec:relatedworks}

Due to the Joint Aerial Layer Networking (JALN) activities of the U.S. Air Force, design of a robust and resilient Airborne Network (AN) has received considerable attention in the networking research community in recent years. It has been investigated that the flat ad hoc networks have limitations with respect to data transmission, distance, interference and scalability \cite{Milner03, Milner09, Gupta00}.  Accordingly, \cite{national1997The, Milner03, Milner09} suggested the addition of a mobile wireless backbone of base stations (analogous to cellular telephony or the Internet backbones), in which topologies and mobility can be controlled for purposes of assured communications. 

There exists considerable amount of studies on topology control using power control in MANETs \cite{Ram00, Ram07, Wu06, Paolo05, Milner09}. The goal of the proposed algorithms is to assign power values to the nodes to keep the network connected while reducing the power consumption. The authors of \cite{Ram00, Ram07} proposed distributed heuristics for power minimization in mobile ad hoc networks and offer no guarantees on the worst case performance. 
Santi in \cite{Paolo05} studied the minimum transmission range required to ensure network connectivity in mobile ad hoc networks. He proved that the critical transmission range for connectivity (CTR) is $c\sqrt{\frac{\ln n}{\pi n}}$ for some constant $c$ where mobility model is obstacle free and nodes are allowed to move only within a certain bounded area. In these studies the mobility patterns are not known unlike the problems studied in this paper where it is assumed that the flight paths of ANPs are predictable. Moreover, this research studies the computation of minimum transmission range in presence of region-based faults and in delay tolerant scenario where it is not the case in previous studies.

In recent times, there has been considerable interest in studying localized, i.e., spatially correlated or region-based faults in networks \cite{Sen06, Sen09, Mod09, Mod10}. In order to capture the notion of locality in measuring the fault-tolerance capability of a network, a new variant of connectivity metric called region-based connectivity
was first introduced by Sen et. al. \cite{Sen06}. 
 Region-based connectivity, for multiple spatially correlated faults, has been studied in \cite{Sen09}. The region-based connectivity
of a network can be informally defined to be the minimum number of nodes that has to fail within any region of the network before it is disconnected.
Neumayer et. al \cite{Mod09} gave an analysis on identifying the most vulnerable parts of the network when the faults are geographically correlated. That is, the analysis gives
locations of disasters that would have the maximum disruptive effect on the network in terms of capacity and connectivity. In \cite{Mod10} Neumayer et. al. evaluates average
two-terminal reliability of a fiber-optic network in polynomial time under the presence of such geographically correlated faults. The networks studied in \cite{Sen06, Sen09,  Mod10, Mod09} are all static. However, ANs under study in this research are dynamic.

There may be times that the backbone network may have to operate in a disconnected mode. 
The last few years have seen considerable interest in the networking research 
community in delay tolerant networks (DTN) design \cite{Fall03}. 
The authors of \cite{Sterbenz02} survey challenges in enhancing the survivability of mobile wireless networks. This paper mentions that one of the aspects that can significantly enhance network survivability is the design for end-to-end communication in environments where the path from source to destination is not wholly available at any given instant of time. In this design adjusting the transmit power of the nodes plays an important role. 
Existing DTN research mainly focuses on routing problem in DTN \cite{Jain04, Alonso03}. The paper \cite{Cao13} provides a survey on routing algorithms for DTN. For the routing algorithms to be effective, every pair of nodes should be able to communicate with each other over time. Therefore, the time evolving DTN should be connected over time. Few papers \cite{Huang11, Huang10} have studied the problem of topology control in DTNs. In these papers, the time evolving network is modeled by space-time graph and it is assumed that the space-time graph is initially connected and the problem is to find the minimum cost connected subgraph of the original graph.  

 These papers have not studied the computation of minimum transmission range of nodes in DTN networks such that the time evolving network is connected over time and to the best of our knowledge, this is the first paper that studies this problem. 
\section{System Model and Architecture}
\label{sec:sysmodel}
In the previous section, we argued that the level of {\em reliability} needed for continuous operation of an AN may be difficult to achieve through {\em a completely mobile, infrastructure-less network} and wherever possible a {\em backbone network} with Airborne Networking Platforms (ANPs) should be formed to enhance reliability. 
 In order to achieve this goal, we propose an architecture of an AN where a set of ANPs form a backbone network and provide reliable communication services to combat aircraft on a mission. In this architecture, the nodes of the backbone networks (ANPs) may be viewed as {\em mobile base stations} with {\em predictable and well-structured flight paths} and the combat aircrafts on a mission as {\em mobile clients}. A schematic diagram of this architecture is shown in Fig.~\ref{fig:airborne}.  In the diagram, the black aircrafts are the ANPs forming the infrastructure of the AN (although in Fig.~\ref{fig:airborne}, only aircrafts are shown as ANPs,  the UAVs and satellites can also be considered as ANPs). We assume that the ANPs follow a circular flight path. The circular flight paths of the ANPs and their coverage area (shaded spheres with ANPs at the center) are also shown in Fig.~\ref{fig:airborne}. Thick dashed lines indicate the communication links between the ANPs.  The figure also shows three fighter aircrafts on a mission passing through space known as {\em air corridor}, where network coverage is provided by ANPs 1 through 5. As the fighter aircrafts move along their flight trajectories, they pass through the coverage area of multiple ANPs and there is a smooth hand-off from one ANP to another when the fighter aircrafts move from the coverage area of one ANP to that of another. At points P1, P2, P3, P4, P5 and P6 on their flight path in Fig.~\ref{fig:airborne}, the fighter aircrafts are connected to the ANPs (4), (2, 4), (2, 3, 4), (3), (1, 3) and (1), respectively.


In this paper, we make a simplifying assumption that two ANPs can communicate with each other whenever the distance between them does not exceed the specified threshold (transmission range of the on board transmitter).  We are well aware of the fact that successful communication between two airborne platforms depends not only on the distance between them, but also on various other factors such as (i) the line of sight between the platforms \cite{TIW08}, (ii) changes in the atmospheric channel conditions due to turbulence, clouds and scattering, (iii) the  banking angle,  the wing obstruction and the dead zone  produced by the wake vortex of the aircraft \cite{EPS04} and (iv) Doppler effect. Moreover, the transmission range of a link is not a constant and is impacted by various factors, such as transmission power, receiver sensitivity, scattering loss over altitude and range, path loss over propagation range, loss due to turbulence and the transmission aperture size \cite{EPS04}. However, the distance between the ANPs remains a very important parameter in determining whether communication between the ANPs can take place, and as the goal of this research is to understand the basic and fundamental issues of designing an AN with twin invariant properties of coverage and connectivity, we feel such simplifying assumptions are necessary and justified. Once the fundamental issues of the problem are well understood, factors (i) - (iv) can be incorporated into the model to obtain a more accurate solution.

For simplicity of analysis, we make two more assumptions. We assume that (i) all ANPs are flying at the same altitude and (ii) they follow a circular flight path. The first assumption allows us to reduce the problem from three dimension to two. However, none of these two assumptions are critical and our analysis technique can easily be extended to scenarios where the ANPs are not flying at the same altitude and they are not following a circular flight path. As a consequence of assumption (i), we can view the $n$ backbone nodes (ANPs) as moving points on a 2 dimensional plane. Let $(x_i(t), y_i(t))$ be the coordinates of the node $i$ at time $t$. The network of flying ANPs gives rise to a {\em dynamic graph} $G(t)=(V, E(t))$ where $V=\lbrace 1,2,\ldots,n\rbrace$ is the set of nodes indexed by the ANPs and $E(t)$ is the set of edges at time $t$. There is an edge between two nodes if their Euclidean distance, $s_{ij}$ is less than the transmission range $Tr$ at time $t$, i.e., $E(t)= \lbrace (i,j)|s_{ij}(t)<Tr\rbrace$. It may be noted that the dynamic graph $G(t)=(V, E(t))$ is completely defined by the following five controlling parameters.

\begin{enumerate}
	\item a set of points $\{c_1,c_2,\ldots,c_n\}$ on a two dimensional plane (representing the centers of circular flight paths),
	\item a set of radii $\{r_1,r_2,\ldots,r_n\}$ representing the radii of circular flight paths,
	\item a set of points $\{p_1, p_2,\ldots, p_n\}$ representing the initial locations of the platforms
	\item a set of velocities $\{v_1,v_2,\ldots,v_n\}$ representing the speeds of the platforms, and
	\item transmission range $Tr$ of the transceivers on the airborne platforms.
\end{enumerate}

In next section we explain the computation of dynamic topology of graph $G(t)=(V, E(t))$ when all five controlling parameters are given.
\section{Dynamic Topology Computation}
\label{sec:dynamictopology}

In this section we answer the following question. Given all five problem parameters including the transmission range of the ANPs, how do you determine if the resulting dynamic graph is connected at all times?

\begin{figure}[!t]
 \centering
 \subfigure[]{
\includegraphics[width=0.42\textwidth,keepaspectratio]{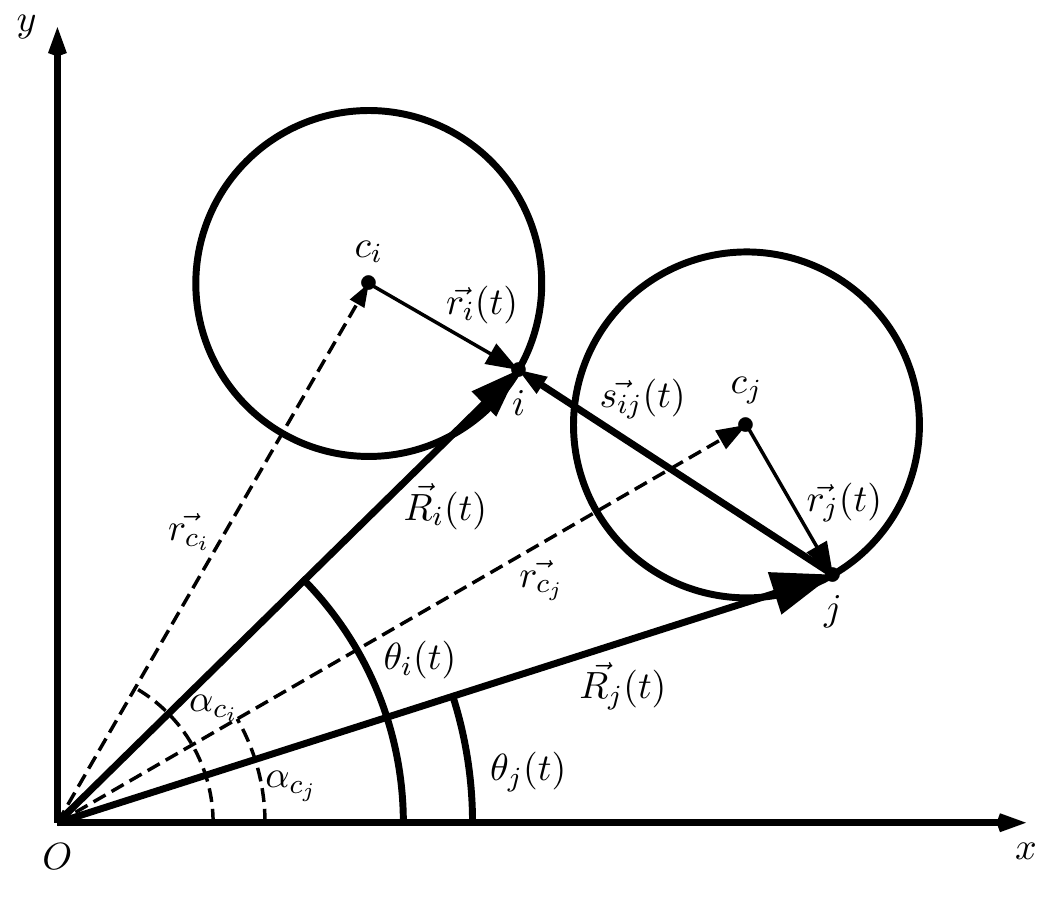}
\label{fig:pavelpolar}}
\subfigure[]{
\includegraphics[width=0.42\textwidth,keepaspectratio]{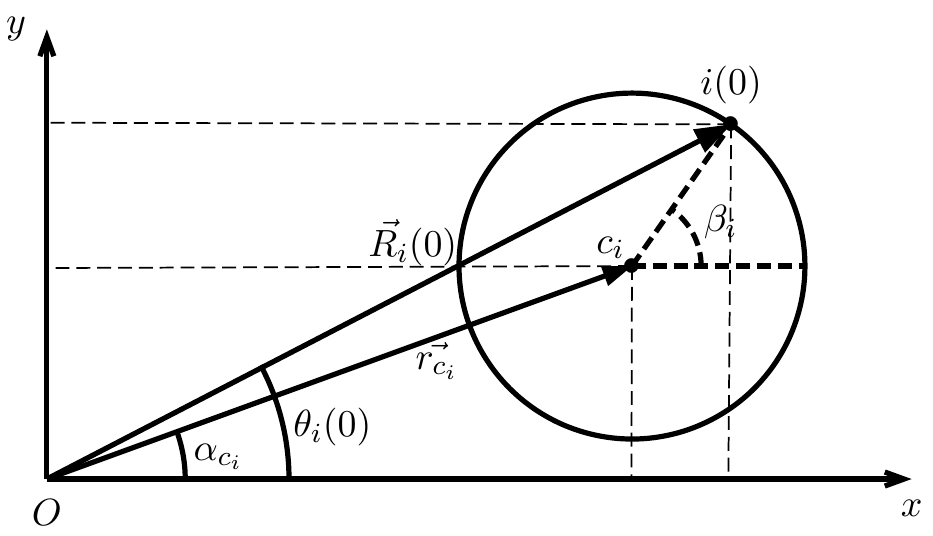}
\label{fig:betafig}}
\caption{(a) Initial phase angle $\beta_{i}$ of point $i$; at time $0$ point is shown as $i(0)$, (b) Vector representations ($\vec{R_{i}}(t)$ and $\vec{R_{j}}(t)$) of two points $i$ and $j$ at time $t$ moving along two circular orbits: $r_{c_{i}}=15,~r_{c_{j}}=27,~\angle{c_{i}Ox}=\alpha_{c_{i}}=\frac{\pi}{3},~\angle{c_{j}Ox}=\alpha_{c_{j}}=\frac{\pi}{6}$}
\end{figure}

Suppose that two ANPs, represented by two points $i$ and $j$ (either in two or in three dimensional space, the two dimensional case corresponds to the scenario where the ANPs are flying at same altitude) are moving along two circular orbits with centers at $c_{i}$ and $c_{j}$ with orbit radius $r_{i}$ and $r_{j}$  as shown in Fig.~\ref{fig:pavelpolar} with velocities $v_{i}$ and $v_{j}$ (with corresponding angular velocities $\omega_{i}$ and $\omega_{j}$), respectively.

A moving  node $i$ is specified by the radius vector $\vec{R_{i}}(t)$ directed from some origin point $O$, and similarly $\vec{R_{j}}(t)$ for point $j$. Therefore the distance $s_{ij}(t)$ between the nodes $i-j$ at  time $t$ is given by:
\begin{equation}
s^{2}_{ij}(t)=(\vec R_{i}(t)-\vec R_{j}(t))^{2}= R_{i}^{2}(t) +  R_{j}^{2}(t) - 2 \vec R_{i}(t) \cdot \vec R_{j}(t)
 \label{eq:vs}
 \end{equation}
As mentioned earlier, we have assumed that the communication between the ANPs is possible if and only if the Euclidean distance between them does not exceed the communication threshold distance $Tr$. This implies that the link between the nodes $i$ and $j$ is alive (or active) when 
\begin{equation}
s_{ij}(t) \leq Tr
\label{eq:sD}
\end{equation}

\medskip
\noindent
In the analysis that follows, we have assumed that ANPs are flying at the same altitude, i.e., we focus our attention to the two dimensional scenario. However, this analysis can easily be extended to the three dimensional case to model the scenario where the  ANPs are flying at different altitude. In this case we can view the ANPs as points on a two-dimensional plane moving along two circular orbits, as shown in Fig. \ref{fig:pavelpolar}.  In Fig.~\ref{fig:pavelpolar}, the vectors from the origin $O$ to the centers of the orbits $c_{i}$ and $c_{j}$ are given as $\vec{r_{c_{i}}}$ and $\vec{r_{c_{j}}}$. The cartesian co-ordinates of the centers can be readily obtained as $\vec{r_{c_{i}}} = (r_{c_{i}}cos~\alpha_{c_{i}}, r_{c_{i}}sin~\alpha_{c_{i}})$ and $\vec{r_{c_{j}}} = (r_{c_{j}}cos~\alpha_{c_{j}}, r_{c_{j}}sin~\alpha_{c_{j}})$. Accordingly, $\vec{R_{i}}(t)$ can be expressed in polar coordinates: $(R_{i}(t) ,\theta_{i}(t))$ with respect to origin point $O$, as shown in Fig.~\ref{fig:pavelpolar}, and similarly for $\vec{R_{j}}(t)$. The initial location of the points $\vec{R_{i}}(0)$ and $\vec{R_{j}}(0)$ are given. From Fig.~\ref{fig:betafig}, the phase angle $\beta_{i}$ for node $i$ with respect to the center of orbit $c_{i}$, can be calculated as (by taking projection on the axes):
\begin{equation}
tan~\beta_{i} = \frac{ R_{i}(0)cos~\theta_{i}(0) - r_{c_{i}}cos~\alpha_{c_{i}} }{ R_{i}(0)sin~\theta_{i}(0) - r_{c_{i}}sin~\alpha_{c_{i}}}
\label{eq:beta}
\end{equation}
From Fig.~\ref{fig:pavelpolar},
\begin{equation}
\vec R_{i}(t)=\vec r_{c_i} + \vec r_{i}(t)
\label{eq:vR}
\end{equation}
where $\vec r_{i}(t) = (r_{i} \cos~(\beta_{i} + \omega_{i}t ), r_{i} \sin~(\beta_{i} + \omega_{i}t ))$ (since angle made by $i$ at time $t$ w.r.t. $c_{i}$ is given by $(\beta_{i} + \omega_{i}t)$). Therefore, the angle between $\vec r_{i}(t)$ and $\vec r_{c_{i}}$ is $(\beta_{i} - \alpha_{c_{i}} + \omega_{i}t )$. Hence, 
\begin{equation}
R_{i}^{2}(t)= r_{c_i}^2 + r_{i}^{2} + 2r_{c_i} r_{i} \cos~(\beta_{i} - \alpha_{c_{i}} + \omega_{i}t )
\label{eq:R}
\end{equation}
Now taking the projection of  $\vec R_{i}(t)=\vec r_{c_i} + \vec r_{i}(t)$ on the $x$ and $y$ axes, we get
\begin{eqnarray}
R_{i}(t)\cos \theta_{i}(t) & = & r_{c_i}\cos~\alpha_{c_i} + r_{i}\cos~(\beta_{i} + \omega_{i}t),
\label{eq:Rx}
\\
R_{i}(t)\sin~\theta_{i}(t) & = & r_{c_i}\sin~\alpha_{c_{i}} + r_{i}\sin~(\beta_{i} + \omega_{i}t)~~~
\label{eq:Ry}
\end{eqnarray}
Recalling $\cos (A-B) = \cos A \cos B + \sin A \sin B$, and simplifying we get
\begin{eqnarray}
R_{i}(t)R_{j}(t)\cos (\theta_{i}(t)-\theta_{j}(t)) & = & r_{c_i} r_{c_j} \cos~\alpha_{c_{i}c_{j}} + r_{i} r_{j} \cos(\beta_{ij}+(\omega_{i}-\omega_{j})t) \nonumber \\
& + & r_{c_i}r_{j} \cos(\alpha_{c_{i}} - \beta_{j} - \omega_{j}t) + r_{c_j}r_{i} \cos(\alpha_{c_{j}} - \beta_{i} - \omega_{i}t)
\label{eq:cosij}
\end{eqnarray}
where $\alpha_{c_{ij}}=\alpha_{c_{i}}-\alpha_{c_{j}}$ and $\beta_{ij} = \beta_{i} - \beta_{j}$. Combining equation \ref{eq:vs} with equations \ref{eq:R} and \ref{eq:cosij}, we have:
\begin{eqnarray}
s_{ij}^{2}(t) & = & r_{c_{i}}^{2} + r_{i}^{2} + 2r_{c_{i}}r_{i}\cos(\beta_{i} - \alpha_{c_{i}}+ \omega_{i}t) + r_{c_{j}}^{2} + r_{j}^{2} + 2r_{c_{j}}r_{j}\cos(\beta_{j} - \alpha_{c_{j}} + \omega_{j}t) \nonumber \\
& - 2[& r_{c_i} r_{c_j} \cos~\alpha_{c_{i}c_{j}} + r_{i} r_{j} \cos(\beta_{ij} +(\omega_{i}-\omega_{j})t)  \nonumber\\
&  & +r_{c_i}r_{j} \cos(\alpha_{c_{i}} - \beta_{j} - \omega_{j}t) + r_{c_j} r_{i} \cos(\alpha_{c_{j}} - \beta_{i} - \omega_{i}t)]
\label{eq:finals1}
\end{eqnarray}

In equation \ref{eq:finals1}, all parameters on the right hand side are known from the initial state of the system, and thus the distance $s_{ij}(t) $ between the nodes  $i-j$ at any time $t$ can be obtained. If the ANPs move at the same velocity, i.e., $\omega_{i}=\omega_{j} = \omega$ for all $i,j$ and the radius of the circular orbits are identical, i.e., $r_{i} = r_{j} = r$ for all $i,j$, and the above expression simplifies to:
\begin{eqnarray}
s_{ij}^{2}(t) &= & r_{c_{i}}^{2} + r^{2} + 2r_{c_{i}}r\cos(\beta_{i} - \alpha_{c_{i}}+ \omega t) + r_{c_{j}}^{2} + r^{2} + 2r_{c_{j}}r\cos(\beta_{j} - \alpha_{c_{j}} + \omega t) \nonumber \\
& -2[ & r_{c_i} r_{c_j} \cos~\alpha_{c_{i}c_{j}} + r^{2} \cos\beta_{ij} \nonumber\\
& & + r_{c_i}r \cos(\alpha_{c_{i}} - \beta_{j} - \omega t) + r_{c_j}r \cos(\alpha_{c_{j}} - \beta_{i} - \omega t)]
\label{eq:finals2}
\end{eqnarray}

\begin{figure*}[!t]
\centering
\subfigure[]{
\includegraphics[width=0.43\textwidth,keepaspectratio]{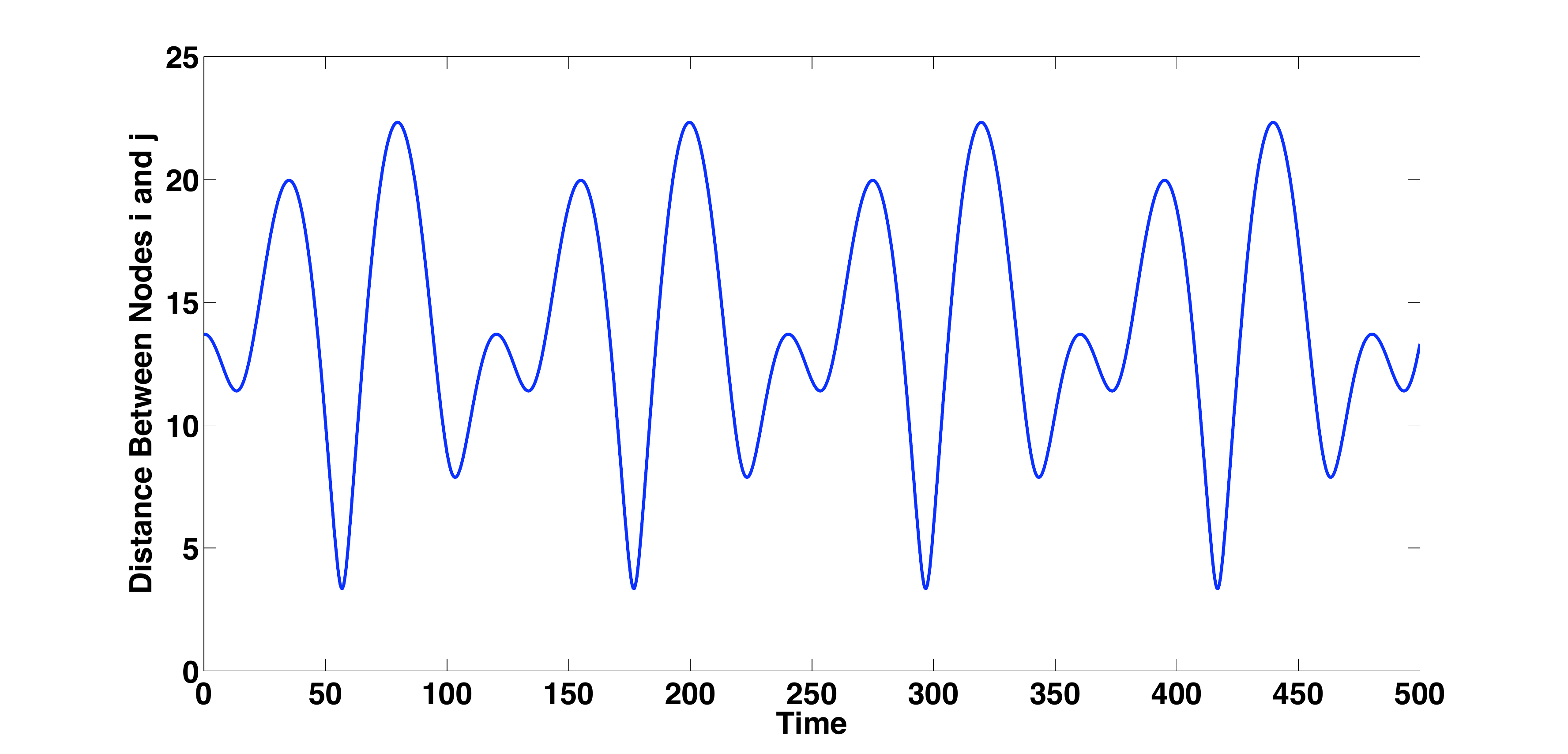}
\label{fig:waveform1}
}
\subfigure[]{
\includegraphics[width=0.48\textwidth,keepaspectratio]{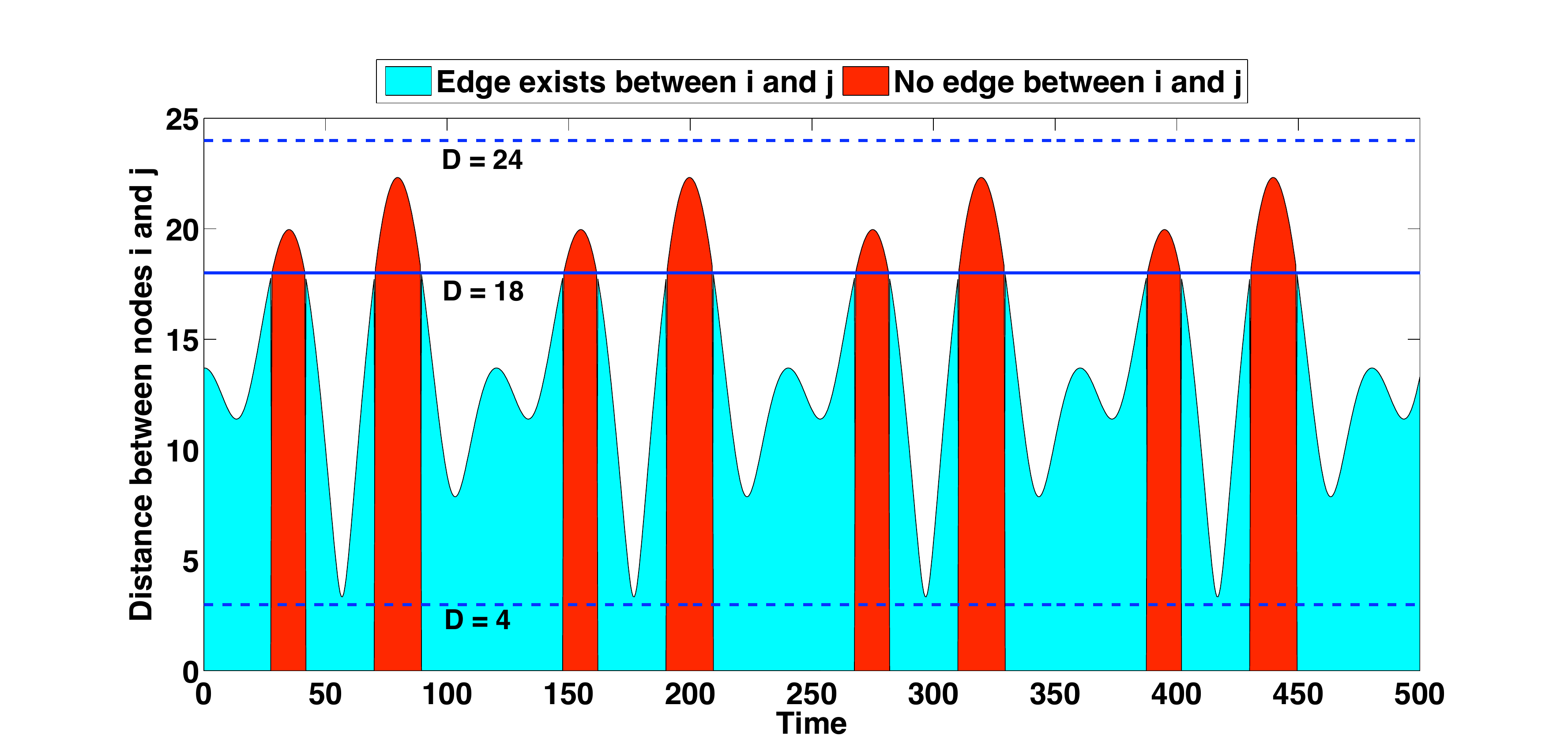}
\label{fig:waveform2}
}
\caption{Effect of the distance between nodes on the existence of the communication link between them; (a)Distance between two points $i$ and $j$ as a function of time, (b)Active (Blue/Light gray)/Inactive (Red/ Dark gray) times of the link between $i$ and $j$ with transmission range $Tr$ = 18}
\label{fig:waveform}
\end{figure*}

An example of a plot of equation (\ref{eq:finals1}) (generated using MATLAB) is shown in Fig.~\ref{fig:waveform1} with  communication threshold distance $Tr =18$. This implies that the link between the nodes $i$ and $j$ exists, when the distance between them is at most $18$ and the link does not exist otherwise. This is shown in Fig.~\ref{fig:waveform2}. The red(dark gray) part indicates the time interval when the link is {\em inactive}(or dead) and the blue(light gray) part indicates when it is {\em active} (or live). 

\begin{figure}[htp]
\centering
\includegraphics[width=0.6\textwidth, keepaspectratio]{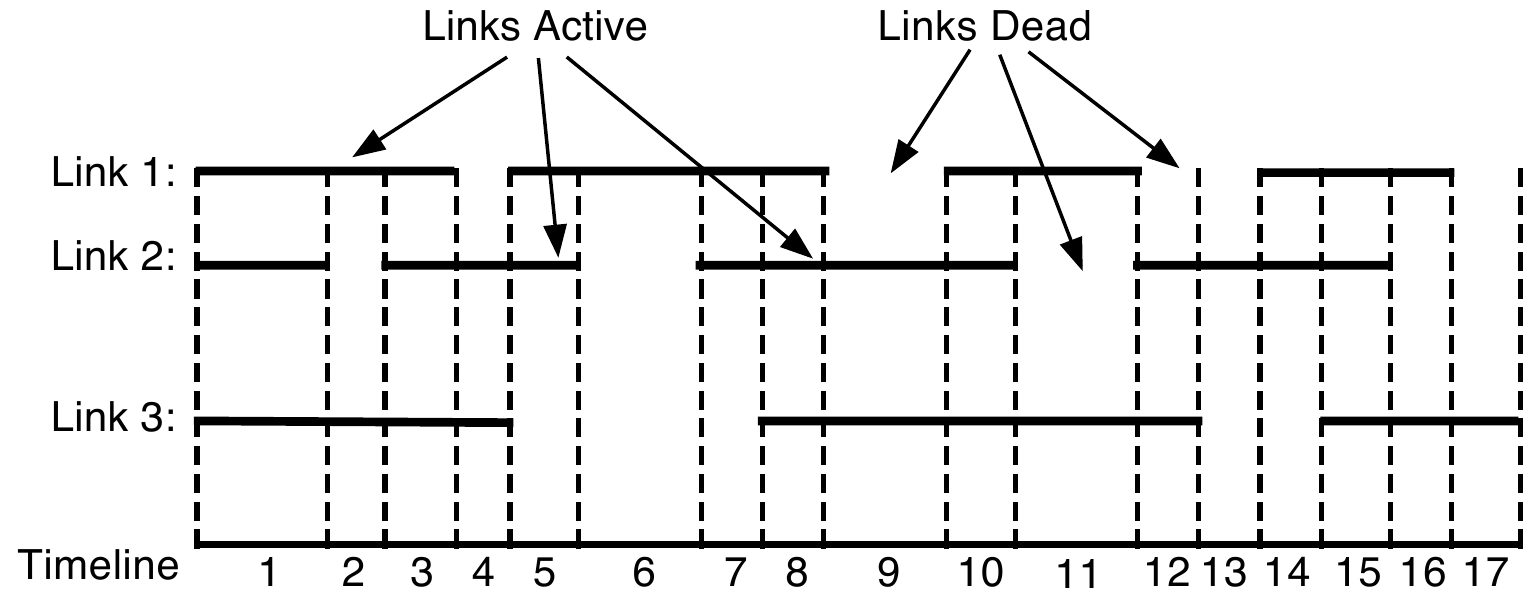}
\caption{Active/Inactive time interval of each link and interval intersection projections on the time line}
\label{fig:interval}
\end{figure}
Thus using equation (\ref{eq:finals1})  and comparing the distance between any two nodes with the communication threshold $Tr$, we can determine active/inactive times of all links. This can be represented as intervals on a time line as shown in Fig.~\ref{fig:interval}.  By drawing projections from the end-points of the active/inactive times of each link on the time line, we can find out all the links that are active during an interval on time line. As shown in Fig.~\ref{fig:interval}, links 1, 2 and 3 are active in interval 1; links 1 and 3 are active in interval 2, links 1, 2 and 3 are active in interval 3 and so on. Once we know all the links that are active during a time interval, we can determine if the graph is connected during that interval using any algorithm for computing graph connectivity \cite{Diestel05}. By checking if the graph is connected at all intervals, we can determine if the graph is connected at all times, when the ANPs are moving at specified velocities.

We note that based on equation (\ref{eq:finals1}), $s_{ij}$ is periodic if every pair of velocities $\omega_i$ and $\omega_j$ are {\em commensurate}, i.e. $\omega_i/\omega_j$ is a rational number \cite{Olmsted72}. 
 Therefore, the network topologies will be repeated periodically and it is enough to check network connectivity in one period.  

If the problem parameters (1) through (5) are specified, we can check if the dynamic graph is connected at all times following these two steps. In the first step, we determine the lifetime (active/inactive intervals) of every link between every pair of nodes $i$ and $j$ by comparing $s_{ij}(t)$ with $Tr$ and finding the time points that the state of a link changes. Let $L(Tr)=\{e_1, e_2,\ldots, e_l\}$ denote the set of events $e_i$s that state of a link changes when transmission range is $Tr$; 
 $L(Tr)$ is sorted in increasing order of the time of the events. Hence, between two consecutive events $e_i$ and $e_{i+1}$ that happen at times $t_i$ and $t_{i+1}$ the set of active links is unchanged.  Algorithm \ref{alg:Link} shows the details of computing  $L(Tr)$. In the second step we check the graph connectivity in each interval $[t_i, t_{i+1})$ for all $0\le i\le l-1$ using connectivity checking algorithm in \cite{Cormen}. $t_0$ shows current time (starting point).  Step 2 is described in detail in Algorithm \ref{alg:conn}.

\begin{algorithm}[H]
\small
\caption{Link Lifetime Computation}
\label{alg:Link}
{\em Input}: (i) a set of points $\{c_1,c_2,\ldots,c_n\}$ representing the centers of circular flight paths,
(ii) a set of radii $\{r_1,r_2,\ldots,r_n\}$ representing the radii of circular flight paths,
(iii) a set of points $\{p_1,p_2,\ldots,p_n\}$ representing the initial locations of the platforms,
(iv) a set of velocities $\{v_1,v_2,\ldots,v_n\}$ representing the speeds of the platforms.\\
{\em Output}: $L(Tr)$: an ordered set of events that the state of a link changes from active to inactive or inactive to active.
\begin{algorithmic}[1]
\STATE $L(Tr)\leftarrow \emptyset$
\FORALL {pairs $i,j$}
\STATE Compute $l$ to be the set of time points $t$ such that $s_{ij}(t) = Tr$ (equation \ref{eq:finals1}) over a period of time, to find the instances of times $t$ where the state of the link $(i,j)$ changes. If $s_{ij}(t) = Tr$ and is $s_{ij}(t)$ increasing at $t$, it implies that the link dies at $t$, and if $s_{ij}(t)$ decreasing at $t$, it implies that the link becomes active at $t$.
\FORALL {$l_k\in l$}
\STATE Find the position of $l_k$ in $L(Tr)$ using binary search and Add the event into $L(Tr)$. ($L(Tr)$ is sorted in increasing order)
\ENDFOR
\ENDFOR
\end{algorithmic}	
\end{algorithm}


\begin{algorithm}[H]
\small
\caption{Checking Connectivity of Airborne Network}
\label{alg:conn}
{\em Input}: $L(Tr)$\\
{\em Output}: {\em true} if graph is connected all the time; otherwise {\em false}.

\begin{algorithmic}[1]
\FORALL {$l_i\in L(Tr)$}
\STATE Check if the AN graph is connected with the set of live links during interval $[l_i,l_{i+1})$. This can be done with the connectivity testing algorithm in \cite{Cormen}
\STATE {\bf if} AN graph is not connected, {\bf return} $false$
\ENDFOR
\STATE {\bf return} $true$
\end{algorithmic}	
\end{algorithm}

Let $n$ be the number of ANPs. The first loop of Algorithm \ref{alg:Link} is executed for $O(n^2)$ times. The number of iterations of the inner loop depends on the number of the solutions of $s_{ij}(t) = Tr$. For the case that ANPs move at the same velocity, i.e., $\omega_i=\omega_j=\omega$ it is obvious that equation (\ref{eq:finals2}) is periodic and length of one periodic interval is $2\pi/\omega$. So, it is enough to execute Algorithm \ref{alg:Link} for one period $[t_0,t_0+2\pi/\omega)$. In this case, equation (\ref{eq:finals2}) can be written as $A\cos(\omega t)+B\sin(\omega t)= \sqrt{A^2+B^2}\sin(\phi+\omega t)$ where $A,B$ and $\phi$ are constants and can easily be obtained from equation (\ref{eq:finals1}). In this case, the equation $s_{ij}(t) = Tr$ can have at most two solutions and the solutions can be found in constant time. Therefore, for every link, the timeline is divided into at most three segments in one period and the size of the set of intervals, $|L(Tr)|$ is $O(n^2)$; also, the time complexity of the binary search is $O(\log n^2)$. So, the total time complexity of Algorithm \ref{alg:Link} is $O(n^2\log n)$. Even when the velocities of the ANPs are different, $s_{ij}$ remains periodic if every pair of velocities $\omega_i$ and $\omega_j$ are {\em commensurate}, i.e. $\omega_i/\omega_j$ is a rational number \cite{Olmsted72}. In this case also we need to solve $s_{ij}(t) = Tr$ for one period only. Otherwise, equation (\ref{eq:finals1}) is not periodic and we need to consider a period of time between $t_0$ and finish time $t_f$ and find the solutions in that period. For the sake of simplicity, in this paper we assume that the ANPs move at the same speed.
The running time of connectivity testing algorithm in \cite{Cormen} is $O(n^2)$. Also, as $|L(Tr)|=O(n^2)$ time complexity of Algorithm \ref{alg:conn} is $O(n^4)$. 

\subsection{Predictable Ill-Structured Flight Path}
\label{sec:illPath}

In this subsection we consider ANPs following predictable ill-structured flight paths. {\em Predictable Ill-structured flight paths} are defined as pre-defined equations in the 3D space (or in 2D space, in case the aircrafts are all moving at the same altitude). Using the same assumption as in the earlier sections, we assume that there exists a communication link between two nodes in such an AN if they are within the specified threshold distance $D$ from each other. Positions of nodes of the network at any time can be found by using the parametric representation as $\vec{r_{i}}(t) = \big(x_{i}(t), y_{i}(t), z_{i}(t)\big)$, for each node $i=1,2,\ldots, N$, where $t$ is the time and $x_{i}(t)$, $y_{i}(t)$, and $z_{i}(t)$ represent the $x$, $y$ and $z$ co-ordinates of the node in the 3D space at time $t$. Using the analysis described in section~\ref{sec:dynamictopology}, we can compute the link lifetimes for all pairs of nodes. Then, using the similar techniques as described in section~\ref{sec:dynamictopology} we can check the connectivity of the dynamic graph formed by the moving ANPs.

\section{Computation of Critical Transmission Range in Fault Free Scenario }
\label{sec:faultfree}

It is conceivable that even if the network topology changes due to movement of the nodes, some underlying structural properties of the network may still remain invariant.
A structural property of prime interest in this context is the {\em connectivity} of the dynamic graph formed by the ANPs. We want the ANPs to fly in such a way, that even though the links between them are established and disestablished over time, the underlying graph remains connected at all times. We define {\em critical transmission range (CTR)} to be the smallest transmission range necessary to ensure network graph $G(t)$ is always connected. We would like to determine $CTR$. In this case, the problem will be specified in the following way. Given controlling parameters 1, 2, 3 and 4, what is the minimum  transmission range of the ANPs so that the resulting graph is {\em connected} at all times? 


In the previous section we explained how we check network connectivity when all five parameters are given. 
The maximum transmission range of an ANP $Tr_{max}$ is known in advance. In order to compute $CTR$ we can conduct a binary search within the range $0 - Tr_{max}$ and we can determine the smallest transmission range that will ensure a connected AN during the entire operational time when all other problem parameters have already been determined. The binary search adds a factor of $\log Tr_{max}$ to the complexity of Algorithms \ref{alg:Link} and \ref{alg:conn}.

\section{Computation of Critical Transmission Range in Faulty Scenario}
\label{sec:faulty}

The CTR computed in previous section may not guarantee the connectivity of backbone network when some of ANPs fail. In this section, we consider the AN scenario where some of the network nodes are faulty and we compute {\em critical transmission range in faulty scenario(CTR$_f$)} which is defined to be the {\em smallest transmission range necessary to ensure network connectivity, irrespective of (a) the location of the fault region and (b) the time of the failure}. First we describe the fault model used in this paper. Also, we identify the challenges that one has to confront, in order to find the {\em CTR$_f$}.

\subsection {Fault Model}
\begin{figure}[htb]
 \center
\includegraphics[width=0.3\textwidth,keepaspectratio]{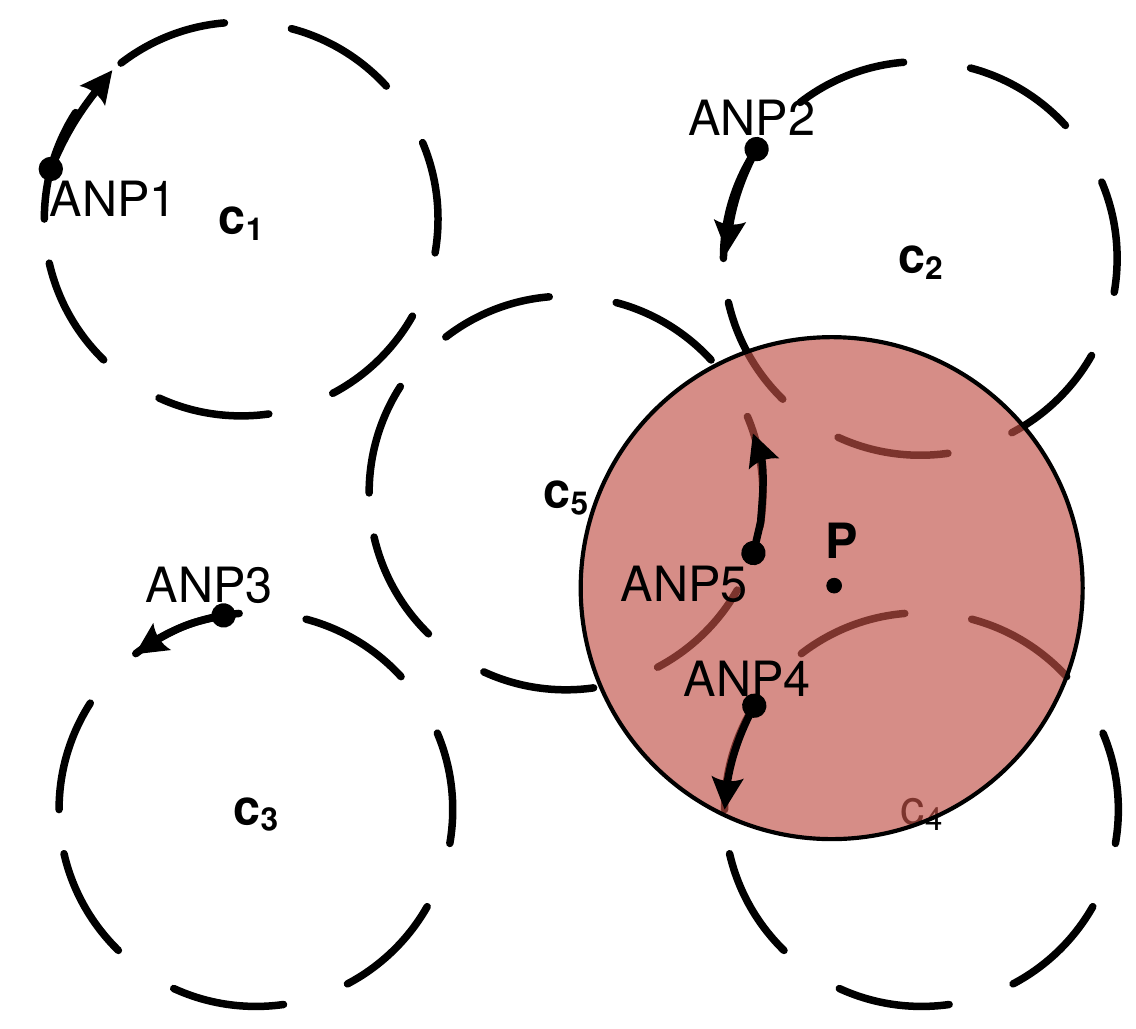}
 \caption{ANPs on a circular flight path on a 2D-plane with a fault region}
\label{faultdiagram}
\end{figure}

As we mentioned before, our focus is on {\em spatially correlated} (or {\em region-based}) faults such as Electromagnetic Pulse (EMP) attacks or jamming. Spatially correlated or region-based faults imply that the faulty nodes due to an attack are confined to a {\em geographic area}. In a two dimensional deployment area, a region can be viewed as a circular area with radius $R$ (in three dimensional space it can be viewed as a sphere with radius $R$). In our model, when a region is under attack and consequently fails at time $t$, some or all the ANPs within that region at time $t$ also fail. In this version of the model, we also make an assumption that only one region can fail at any one time.  Fig. \ref{faultdiagram} shows five ANPs moving on a two dimensional plane and a faulty region (red circle, centered at point $P$) at time $t$. Since ANPs 4 and 5 are within the fault region at time $t$, we assume that these nodes are damaged and no longer can be viewed as part of the backbone network. 
  It may be noted that both the location of the center of the fault circle, $P$, as well as the time of attack, $t$, play a critical role in determining the impact of the attack on the backbone network.

\subsection {Problem Formulation and Design Challenges}

In faulty scenario the connectivity problem is defined in the following way: Given the controlling parameters 1, 2, 3 and 4 (defined in Section \ref{sec:sysmodel}) as well as the radius of a region $R$, what is the {\em smallest transmission range necessary to ensure network connectivity, irrespective of (a) the location of the fault region and (b) the time of the failure}. In other words, the problem is how to compute CTR$_f$. 

One can easily recognize the complexity of the problem by noting that potentially there could be an infinite number of locations for point $P$ and infinite choices for attack time $t$. In our analysis we show that although there could be an infinite number of choices of $P$ and $t$, we need to consider only a small subset of them to correctly determine CTR$_f$. 
 The tasks that need to be performed before a solution to the problem is found can be listed as follows:

\begin {itemize}
\item Computation and comprehension of the dynamic topology of the backbone network (in a fault-free scenario) as it changes with movements of the ANPs.
\item How many regions (locations of point $P$) and instances of attack time $t$ should be considered?
\item How to determine the ANPs that are damaged when an attack takes place in location $P$ at time $t$?
\end{itemize}
In section \ref{sec:dynamictopology} we described the computation of the dynamic topology of the backbone network (in a fault-free scenario). In the following subsections we describe our techniques to deal with the second and third challenges and to compute CTR$_f$.

\subsection{Regions to Examine}
\label{regions}
\begin{figure}
\centering
\includegraphics[width=0.4\textwidth, keepaspectratio]{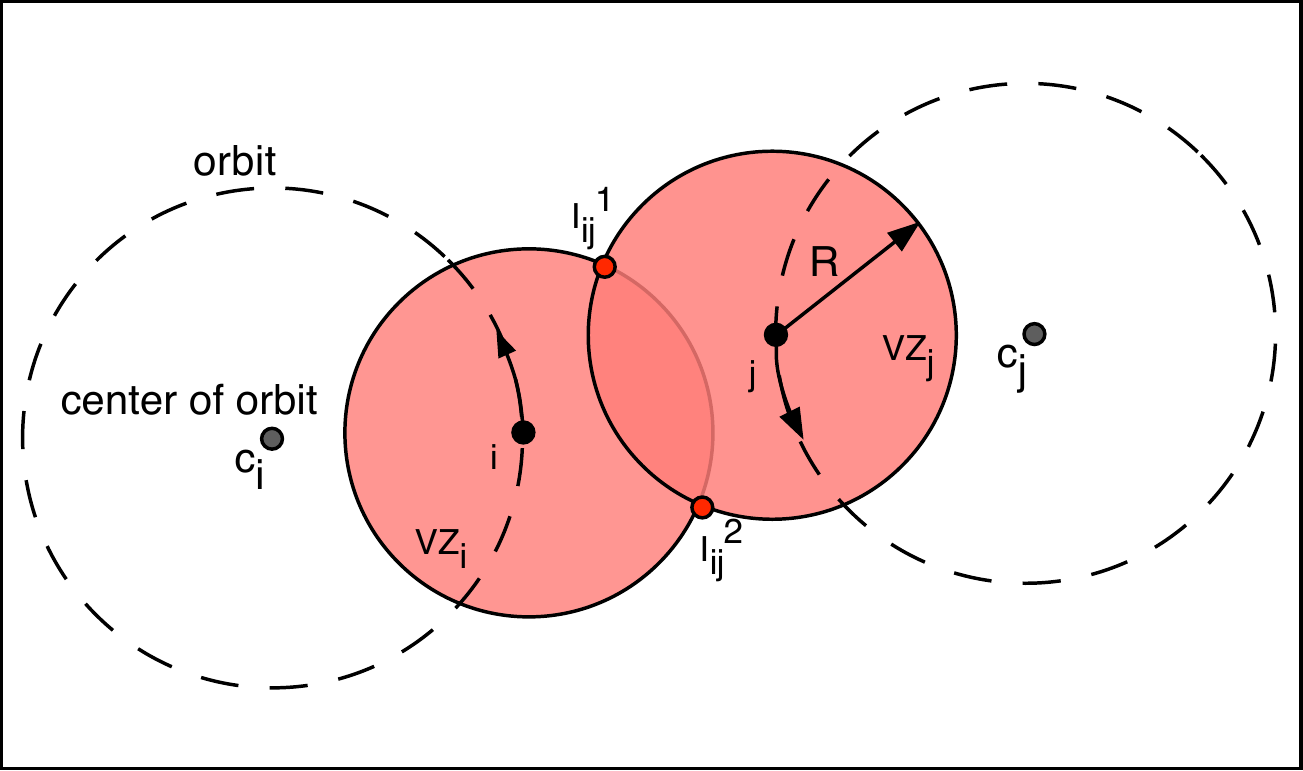}
  \caption{$I_{ij}^1$ and $I_{ij}^2$ are intersection points of $VZ_i$ and $VZ_j$ at time $t$.}
  \label{fig:Ipoint}
\end{figure}

The authors in \cite{Sen06} introduced the notion of {\em region-based} faults and introduced a new metric, {\em region-based connectivity}, to measure the fault-tolerance capability of a network under the region-based fault model. Region-based connectivity of a network is defined to be the minimum number of nodes that has to fail in any region of the network before it is disconnected. In this study, a region is defined to be a circle of radius $R$. With this definition of a region, the number of potential regions could be infinite. The authors in \cite{Sen06} proved that in a static wireless network, only a limited number of distinct regions need to be examined to compute the region-based connectivity. They showed that it is enough to consider the regions centered at the intersection points of the circles centered at the nodes with radius $R$. Although the AN is dynamic, if we take a snapshot of the network at some instance of time $t$, the AN can be viewed as a static network with a specific topology and nodes in specific locations on the plane. The vulnerability zone of a node $i$, $VZ_i(t)$, is defined to be a circular region centered at the location of node $i$ at time $t$ with radius $R$. The motivation for this definition of the vulnerability zone of node $i$ is the following. If the center of the fault region is within the vulnerability zone of ANP$_i$ (node $i$), then the ANP$_i$ is likely to be damaged. The vulnerability zones of ANPs are shown in Fig.~\ref{fig:Ipoint}. Since there is no discernible difference between a static sensor network considered in \cite{Sen06} and a snapshot of an AN at a specific instance of time $t$, using the analysis presented in \cite{Sen06}, we can conclude that it is enough to examine only the regions centered at the intersection points (I-points) of the vulnerability zones of the ANPs. The vulnerability zones of two ANPs and their intersection points are shown in Fig.~\ref{fig:Ipoint}. If a $VZ_i$ does not have intersection with any other node's vulnerability zone, an I-point is considered at the location of the node $i$.

Since the ANPs are mobile, the location of the intersection points of their vulnerability zones also changes with time. Each pair of ANPs will have at most two intersection points. Since there are only $n(n-1)/2$ pairs of nodes, at most $n(n-1)$ intersections points can exist at any given time (it may be noted that depending on flight path of a pair of ANPs, their vulnerability zones may never intersect). We define a set of $n(n-1)+n$ I-points, $\mathcal{I}=\{I_{(1,2)}^1, I_{(1,2)}^2,I_{(1,3)}^1, I_{(1,3)}^2,\ldots, I_{(n-1,n)}^1,I_{(n-1,n)}^2, I_1, I_2,\ldots,I_n\}$, where the $I_{(i,j)}^1$ and $I_{(i,j)}^2$ are the intersection points of the vulnerability zones $VZ_i$ and $VZ_j$. We will use the notation $I_{(i,j)}^1(t)$ and  $I_{(i,j)}^2(t)$ to denote the locations of $I_{(i,j)}^1$ and $I_{(i,j)}^2$ at time $t$.  Similarly, $I_i(t)$ will denote the location of node $i$ at time $t$. Based on the results presented in \cite{Sen06}, it is known that at any point of time $t$ it is sufficient to examine only the regions centered at the I-points in $\mathcal{I}$. In the rest of the paper we will use $I_{(i,j)}$ to denote both $I_{(i,j)}^1$ or $I_{(i,j)}^2$.

For every two nodes $i$ and $j$, $VZ_i(t)$ and $VZ_j(t)$ intersect iff 
 $s_{ij}(t)\leq 2R$. In this case we say that the {\em region} centered at intersection point $I_{(i,j)}$ {\em exists} at time $t$; otherwise, it does not, i.e., there exists no region that can cover both nodes $i$ and $j$ at time $t$. It may be noted that due to the mobility of the ANPs, $I_{(i,j)}$ may exist at some point of time $t$ and may not exist at some other point of time $t'$. By checking the condition $s_{ij}(t)\leq 2R$, we can determine the intervals on the timeline when  $I_{(i,j)}$ exists 
  for each pair of nodes $i$ and $j$; i.e., we can compute {\em existence} intervals of each I-point on the timeline.  Let $T(f)=\{(t_{f}^1,t_{f}^2),\ldots,(t_{f}^{k-1},t_{f}^k)\}$ be the set of existence intervals of I-point $f\in \mathcal{I}$ where the first element in every pair $(t_{f}^j,t_{f}^{j+1})$ is the start time and the second one is the finish time of the $j$- th existence interval. If in a time interval $(t_{I_i}^j,t_{I_i}^{j+1})$, $VZ_i$ does not have intersection with any other ANP's vulnerability zone then a region centered at $I_i$ should be considered, i.e, $(t_{I_i}^j,t_{I_i}^{j+1})\in T(I_i)$. Without loss of generality we can assume that the region centered at the point $I_i$ exists all the time and it only covers node $i$. 
The computation of the intervals on the timeline when  $I_{(i,j)}$ exists (or does not exist), for each pair of nodes $i$ and $j$, can be carried out
by an algorithm similar to Alg. \ref{alg:Link} presented earlier. The only differences are (i) the value of $t$ that satisfies the equation $s_{ij}(t) = 2R$ should be computed instead of the value of $t$ that satisfies the equation $s_{ij}(t) = Tr$, (ii) since there is no need to combine existence interval information of one pair of nodes ($I_{(i,j)}$) with another pair, the binary search in step 5 of Alg. \ref{alg:Link} is not needed.

\subsection{Computation of the Damaged ANPs in a Fault Region }
After finding the existence intervals of I-points we want to find the set of nodes that might be damaged by the failure of a region centered at an I-point when it exists. A node might be damaged by failure of a region if the Euclidean distance between the center of the region and the node is less than $R$. As explained in part \ref{regions} the regions centered at I-point $I_i \in \mathcal{I}$s only can destroy node $i$. Since, we know the locations of each pair of nodes $i$ and $j$ at time $t$, we can compute $VZ_i(t)$ and $VZ_j(t)$, 
 and hence $I_{(i,j)}^1(t)$ and  $I_{(i,j)}^2(t)$, the intersection points of $VZ_i(t)$ and $VZ_j(t)$ at time $t$.

\begin{figure}[htb]%
\centering
\includegraphics[width=0.4\textwidth]{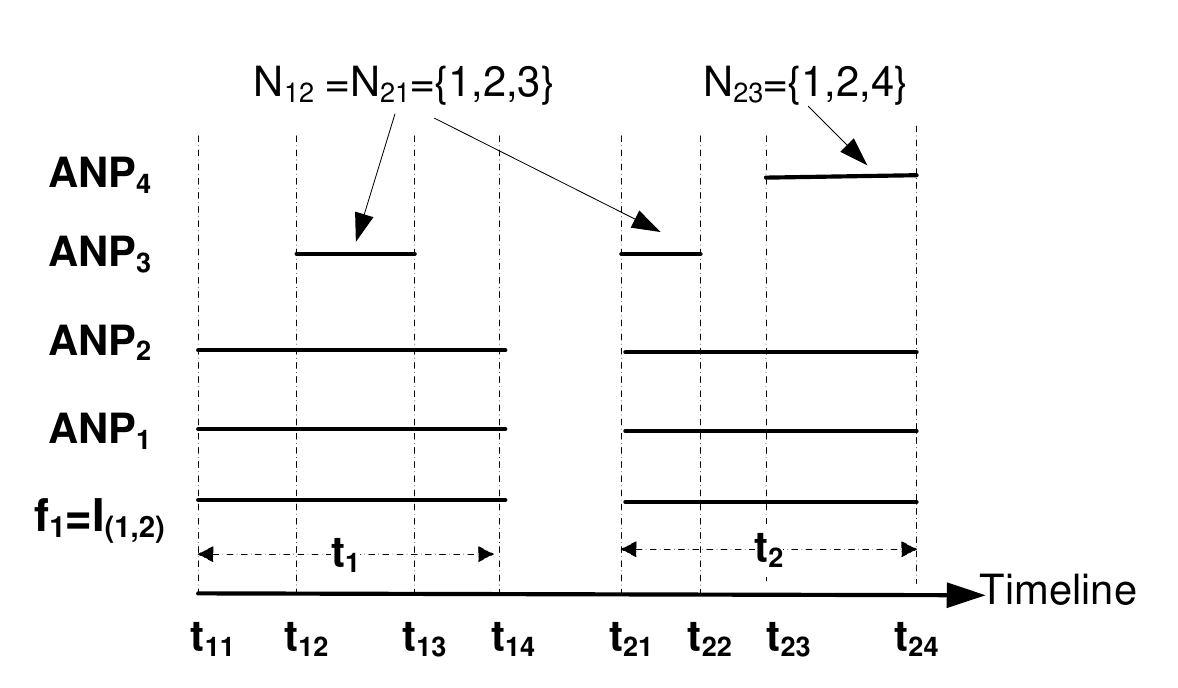}%
\caption{Region coverage timeline of the region centered at $f_1=I_{(1,2)}$; The first timeline shows the availability intervals of $f_1$; i.e, $T(f_1)=\{t_1,t_2\}$. }%
\label{nodelife}%
\end{figure}

Once the location of each intersection $I_{(i,j)}$ in its existence intervals $\in T(I_{(i,j)})$ are known, we can find the nodes that might be damaged if the region centered at $I_{(i,j)}$ fails. For ease of notation we denote the set of I-points $\in \mathcal{I}$ as $F=\{f_1, f_2,\ldots, f_l\}$. $D_{ik}(t)$ denotes the distance between 
I-point $f_i\in F$ and node $k$ at time $t$. For every I-point $f_i\in F$ 
in its existence interval $\in T(f_i)$, we find $D_{ik}(t)$ for all $k\in V$. Since we know the position of the nodes and I-points at any point of time, $D_{ik}(t)$ can be computed easily. If $D_{ik}(t)\leq R$, then the node $k$ may be damaged due to the region failure $f_i$. From this calculation, we can find out the {\em time interval} when node $k$ is vulnerable to a failure $f_i$. In other words, we can find out the time intervals when a node $k$ is {\em covered} by the region centered at $f_i$ (i.e., $D_{ik}(t)\leq R$). It may be noted that this time interval will be subinterval of the intersection points existence time interval. Accordingly, every existence interval $(t_{f_i}^j,t_{f_i}^{j+1})\in T(f_i)$ is divided into a set of smaller subintervals such that each of these intervals identify a specific set of nodes that may be damaged if the region centered at $f_i$ fails. Suppose that  $t_m$ be the $m$th interval of $T(f_i)$.  We define a set $NT(f_i,t_m)=\{(t_{m1},N_{m1}), (t_{m2},N_{m2}),\ldots, (t_{mj},N_{mj})\}$as the set of subintervals into which $t_m$ is divided, where $t_{mj}$ denotes the start time of the $j$th subinterval of $t_m$ where at least a node enters the region or leaves the region and $N_{mj}$ denotes the set of nodes within the region centered at $f_i$ in its $j$th subinterval. We need to compute $NT(f_i,t_m)$ for every region $f_i\in F$ and for all of its existence intervals.  Based on $NT(f_i,t_m)$ we can draw a timeline, {\em region-coverage timeline} for each region centered at an I-point $f_i\in F$. Fig. \ref{nodelife} shows an example in which $NT(f_1,t_1)=\{(t_{11},\{1,2\}), (t_{12},\{1,2,3\}), (t_{13},\{1,2\})\}$. 

\subsection{Computation of Critical Transmission Range in Faulty Scenario ({\em CTR$_f$})}
\label{Algorithm}
\vspace{-0.05in}
In this section we propose an algorithm to find {\em CTR$_f$}. 
 
The transmission range $Tr$ is one of the parameters that determines the number of active links at any given time. Similarly, the location of the center of the fault region is one of the parameters that determines the number of ANPs that can potentially be damaged by the fault. For a specific region centered at an I-point $f_i$, and a transmission range $Tr$, we define an interval on the timeline as {\em static interval}, if the set of potentially damaged nodes due to a  region fault at location $f_i$ and the set of alive links with transmission range $Tr$ remain unchanged.
We can find static intervals using the timeline  {\em region-coverage($f_i$)} and the timeline {\em link-lifetime} $L(Tr)$. In order to find the static intervals for I-point $f_i$ and transmission range $Tr$, we define four events during the time interval when $f_i$ {\em exists}: (i) a dead link comes alive, (ii) a live link dies, (iii) an ANP node comes within coverage area of $f_i$ and (iv) an ANP node moves out of the coverage area of $f_i$. The instance of time at which any of the four events takes place is the instance of the start time of a  new static interval. Let $SI(f_i, Tr)$ be a sorted list of events resulting from combining the sorted list $L(Tr)$ and $NT(f_i,t_m)$ for all $t_m \in T(f_i)$. Therefore, between any two consecutive elements in $SI(f_i, Tr)$ neither the topology nor the region coverage changes.
 
Once the nodes within a region (or nodes covered by a region) and the set of active links during a static interval are known,  we can use Algorithm-2 in \cite{Sen06} in order to find the {\em region based connectivity} of the network with respect to I-point $f_i$. Region based connectivity with respect to an I-point $f_i$ ($RBC(f_i)$) is defined to be the minimum number of nodes in the region centered at  $f_i$ whose failure disconnects the network. If the number of nodes that can be damaged due to a region based fault at $f_i$ is $n_i$ (i.e., the fault at $f_i$ covers $n_i$ nodes), we would like the ANPs to have enough transmission range, so that the region based connectivity of the graph is at least  $n_i+1$. This will ensure that the network will remain connected if any subset of the covered nodes fails. Using Algorithm-2 of \cite{Sen06}, and applying binary search within the range $0-Tr_{max}$ we can find the minimum transmission range necessary in each static interval, to ensure that the network remains connected when a region $f_i$ fails (during the interval when it $f_i$ is exists). We define $err$ to be the maximum acceptable difference between the smallest transmission range necessary to maintain connectivity and the smallest transmission range computed by the algorithm to maintain connectivity. In our algorithm, we set the maximum possible transmission range to be equal to diameter of the deployment area. The algorithm computes the {\em minimum} transmission range necessary to maintain connectivity for each static interval. The {\em maximum} of these {\em minimum values} computed is the {\em critical transmission range (CTR$_f$)
}. Alg. \ref{alg:minTr} provides all the details.

%
%
 
%
%
 
%
%
 
In Alg. \ref{alg:minTr}, line 1 takes $O(n^2)$. In order to compute $T(f_i)$ we need to solve $s_{ij}=2R$. As described in Alg. \ref{alg:Link}, for the case that ANPs move at the same velocity, $\omega$, this equation can be solved easily in constant time and it has two solutions in one period. So, $|T(f_i)|\leq 2$. In line 5, we have to solve $D_{iv}(t)=R$ for all $v\in V$. For one node $v$, this equation also in one period can have a constant number of solutions since it can easily be converted to a single variable polynomial equation with degree 6. So, computation of $NT(f_i,t_m)$ takes $O(n)$ and $\sum_{t_m}{|NT(f_i,t_m)|}=O(n)$. Consequently, lines 2-7 have complexity of $O(n^3)$. The {\em while} loop is repeated for $\log Tr_{max}$ (binary search complexity). As it is discussed in Alg. \ref{alg:Link}, computation of $L(Tr)$ takes $O(n^2\log n)$. Computation of $SI(f_i, Tr)$ need sorting the sorted lists $NT(f_i,t_m)$ and $L(Tr)$ which takes $O(n^2)$. Clearly, $|SI(f_i, Tr)|=O(n^2)$. Computing $RBC(f_i)$ takes $O(n^4)$  \cite{Sen06}. Therefore, the time complexity of Alg. \ref{alg:minTr} is $O(n^8\log Tr_{max})$.
\begin{algorithm}[tbh]
\small
\caption{Computing $CTR_f$}
\label{alg:minTr}
\begin{algorithmic}[1]
\STATE Compute $s_{ij}(t)$ for all pair of ANPs $i$ and $j$
\STATE \textbf{for all} {I-points $f_i\in\mathcal{I}$}
\STATE \ \ \ Compute 
$T(f_i)=\{(t_{f}^1,t_{f}^2),\ldots,(t_{f}^{k-1},t_{f}^k)\}$
\STATE \ \ \ \textbf{for all} {$t_m\in T(f_i)$}
\STATE \ \ \ \ \ \ Compute \footnotesize $NT(f_i,t_m)=\{(t_{m_1},N_{m_1})
,\ldots,(t_{m_p},N_{m_p})\}$ 
\STATE $error=Tr_{max}$, $tr_a=0$, $tr_b=Tr_{min}=Tr_{max}$
\STATE \textbf{while} {$error>err$}
\STATE \ \ \ $error=error/2$, $Tr=(tr_a+tr_b)/2$
\STATE \ \ \ Find $L(Tr)=\{e_1, e_2,\ldots, e_t\}$ using Alg. \ref{alg:Link} 
\STATE \ \ \ \textbf{for all} {I-points $f_i\in\mathcal{I}$}
\STATE \ \ \ \ \ \ $SI(f_i, Tr)$ $\leftarrow$ Sort the lists $NT(f_i,t_m)$ and $L(Tr)$ based on time\\ \ \ \ \ \ \ of the events (considering all $t_m\in T(f_i)$)
\STATE \ \ \ \ \ \ \textbf{for all} {$event\in SI(f_i, Tr)$}
\STATE \ \ \ \ \ \ \ \ \ Update the graph $G(t)$ (by adding or removing the links) or the\\ \ \ \ \ \ \ \ \ \ region coverage of $f_i$
\STATE \ \ \ \ \ \ \ \ \ $RBC(f_i)\leftarrow$ Using Alg. 2 in \cite{Sen06} Compute the region-based\\ \ \ \ \ \ \ \ \ \ connectivity considering only one region centered at I-point $f_i$
\STATE \ \ \ \ \ \ \ \ \ \textbf{if} ($RBC(f_i)\geq n_i+1$)  NextSI $\leftarrow$ TRUE
\STATE \ \ \ \ \ \ \ \ \ \textbf{else} NextSI $\leftarrow$ FALSE;  break;
\STATE \ \ \ \ \ \ \textbf{if} (NextSI $=$ FALSE)  $tr_a=Tr$; break;
\STATE \ \ \ \textbf{if} (NextSI $=$ TRUE) $tr_b=Tr$; $Tr_{min}=Tr$
\RETURN $Tr_{min}$
\end{algorithmic}
\end{algorithm}


\section{Computation of Critical Transmission Range in Delay Tolerant Airborne Networks {\em CTR$_D$}}
\label{sec:DTN}
In previous sections we explained the computation of critical transmission range in fault free (CTR) and faulty scenarios (CTR$_f$). However, it may not be possible to equip the ANPs with radios that have coverage of radius CTR. Therefore, the backbone network cannot be connected all the times. On the other hand, based on the type of data that should be transmitted between ANPs, data transmissions may be tolerant to some amount of delay. Hence, ANPs may not be needed to have end-to-end paths all the times but they should be able to transmit data to each other in some limited time through intermediate nodes in different network topologies. In this section we investigate the problem of computation of minimum transmission range in delay tolerant airborne networks.

We consider that the trajectories and the distance function $s_{ij}(t)$ of the nodes are periodic over time. As a consequence, the network topologies are repeated periodically. However, periodicity is not an underlying assumption and our results can be utilized in non-periodic scenario as long as the node trajectories for the whole operational duration of a network are given. In Section \ref{sec:dynamictopology} we explained how we can compute {\em link lifetime timeline} and accordingly the network topologies caused by ANPs mobility in a time period when all five controlling parameters are given. We represent the set of topologies in a periodic cycle starting from time $t_0$ (starting time of network operation) by the set $\mathcal{G} =\{G_1, G_2, \ldots, G_l\}$. Each network topology $G_i$ exists for a time duration of $T_i$. As the focus of this section is study of the delay caused by network disconnection (which may be viewed as delay due to queuing at an intermediate node), we assume that other delays due to transmission and propagation are negligible. 

\begin{figure}[htb]
 \center
\includegraphics[width=0.45\textwidth,keepaspectratio]{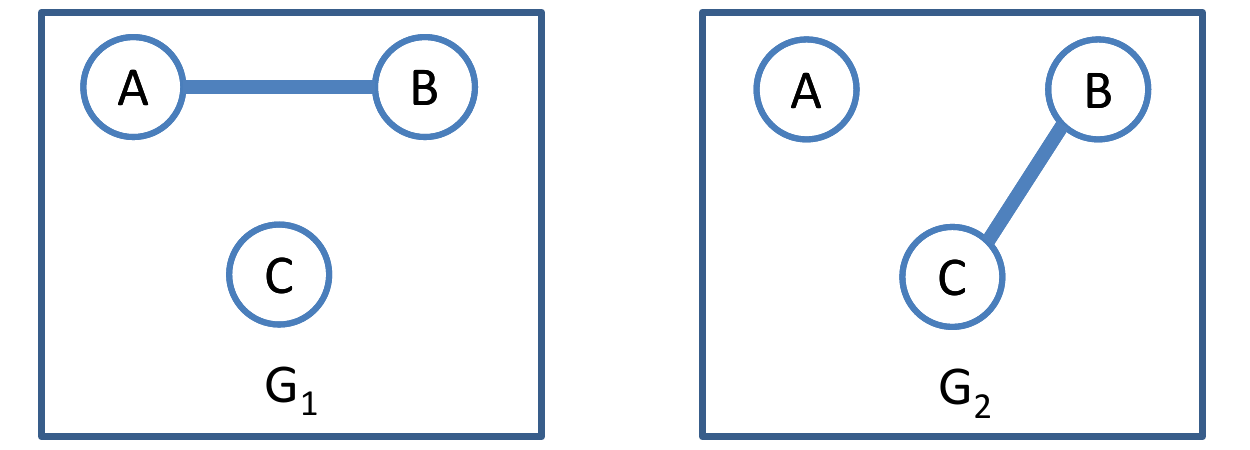}
 \caption{A dynamic graph with two topologies $G_1$ and $G_2$}
\label{fig:DTNex}
\vspace{-0.1 in}
\end{figure}

In Fig. \ref{fig:DTNex}, an example of a dynamic graph with two topologies $G_1$ and $G_2$ in one periodic cycle is shown. $G_1$ and $G_2$ last for $T_1$ and $T_2$ time units respectively. It can be observed that there is no end-to-end path from $A$ to $C$ in either $G_1$ or $G_2$. However,   $A$ can transmit data to $B$ in $G_1$, and $B$ can forward it to $C$ in $G_2$. In this case we say that $A$ can reach $C$ through a temporal path with delay equal to the lifetime of $G_1$, i.e. $T_1$; and the temporal path is completed in $G_2$. We define a {\em temporal path} from node $s$ to $d$ to be a set of tuples $\{(t_1,(v_1,v_2)), (t_2,(v_2,v_3)),\ldots, (t_{k},(v_{k}, v_{k+1}))\}$ such that $v_1=s, v_{k+1}=d$, $v_i\in V$ and for every tuple  $(t_i,(v_i,v_{i+1}))$ , edge $(v_i,v_{i+1})$ is active at time $t_{i}$, and  $t_i\ge t_{i-1}$ for all $1\le i\le k$. Moreover, without loss of generality, we assume that $t_i$ corresponds to the starting time of a topology in $\mathcal{G}$. Then the path delay is defined to be $t_k-t_0$ where $t_0$ 
 is the starting time of $G_1$ in the first periodic cycle. We note that all path delays are computed with respect to starting point $t_0$ but we later show that we can modify the starting point to any time.

We note that existence of a path from node $i$ to $j$ with some delay does not guarantee the existence of a path from $j$ to $i$ with the same delay. For example, in Fig. \ref{fig:DTNex} the path from $C$ to $A$ has a delay of $T_1 + T_2$ while the path delay from $A$ to $C$ is equal to $T_1$. We say that a dynamic graph $G(t)$ is {\em connected with delay $D$} if there exists a temporal path from every node $i\in V$ to every node $j\in V-\{i\}$ with delay smaller than $D$. In a network, if the transmission range $Tr$ is too small, ANPs may not be able to reach each other at all; i.e. there is no temporal path of finite delay between the ANPs. We define {\em critical transmission range in delay tolerant network (CTR$_D$)} to be the minimum transmission range necessary to ensure that the dynamic graph is {\em connected with delay $D$}. 
 We define {\em the connectivity problem in delay tolerant networks} as the problem of computation of CTR$_D$ given the first four controlling parameters defined in Section \ref{sec:sysmodel}, and the delay threshold $D$. 

In order to find the value of CTR$_D$, first we explain an algorithm to check whether a transmission range $Tr$ is adequate for having a connected dynamic network with delay $D$. Using Algorithm \ref{alg:Link} in Section \ref{sec:dynamictopology} we can compute the different network topologies and their lifetime in one periodic cycle. 
 Before describing the rest of the algorithm, first we propose an observation. 

\begin{observation}     
For a given transmission range $Tr$, there is a temporal path from every node $u$ to every node $v$ with finite delay iff the superimposed graph $G_c = \{V, \bigcup_{i=1}^l E_i\}$, where $E_i$ is the set of edges in $G_i$, is connected.   
\end{observation}

\begin{figure}[htb]
 \center
\subfigure{\includegraphics[width=0.2\textwidth,keepaspectratio]{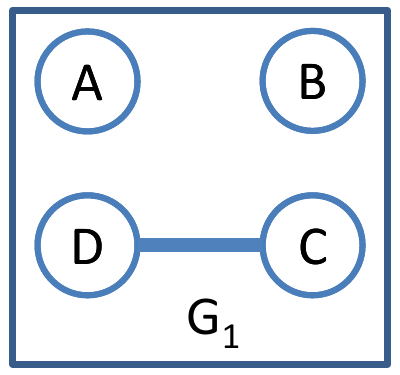}
\label{fig:DTNdelay1}}
\subfigure{\includegraphics[width=0.2\textwidth,keepaspectratio]{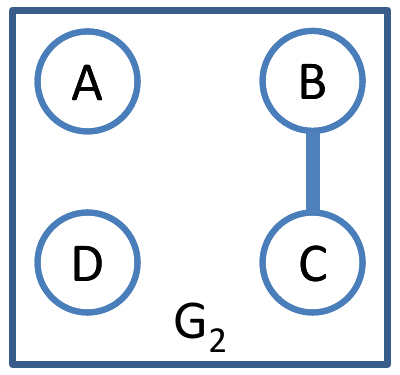}
\label{fig:DTNdelay2}}
\subfigure{\includegraphics[width=0.2\textwidth,keepaspectratio]{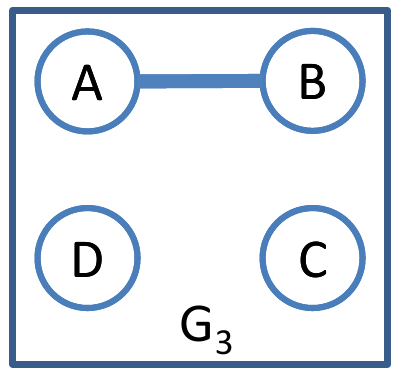}
\label{fig:DTNdelay3}}
 \caption{A dynamic graph with three topologies $G_1$, $G_2$ and $G_3$}
\label{fig:DTNdelay}
\end{figure}

Although a transmission range $Tr$ may be enough to result in a connected superimposed graph $G_c$, it may not be sufficient for the existence of a temporal path between every pair of nodes with delay smaller than a threshold $D$ even if $D$ is as large as $\sum_{i=1}^{l-1}{T_i}$. Fig. \ref{fig:DTNdelay} depicts an AN with three topologies in one period. It can be observed that $A$ cannot have a temporal path from $A$ to $D$ in the first period. Actually the fastest path includes edges $(A,B)$ in $G_3$ in first period,  $(B,C)$ in $G_2$ in the second period and $(C,D)$ in $G_1$ in the third period. Therefore, the path delay is $2(T_1+T_2+T_3)$. Generally, in the worst case in every period just a subpath (a set of consecutive edges) in one topology is used and therefore the maximum delay of a temporal path will be $D_{max}=(l-1)\sum_{i=1}^{l}{T_i}$. Hence, if $D\ge D_{max}$, examining the connectivity of $G_c$ is enough to decide whether for a transmission range there exists a temporal path of delay smaller than $D$ between every pair of nodes in the dynamic network. 

\vspace{0.5 in}
 Next, we explain the algorithm that checks for a given value of transmission range $Tr$ whether a network is connected with delay $D$ where $D < D_{max}$. 
Let $N(u)$ denotes the set of nodes that are reachable from $u\in V$ with delay smaller than $D$. Initially $N(u) = \{u\}$. The algorithm starts by computing the connected components in every topology $G_i$. Let $C_i=\{C_{i,1}, C_{i,2},\ldots C_{i,q_{i}}\}$ represents the set of connected components in $G_i$ where $C_{i,j}$ is the set of nodes in $j$th component of $G_i$ and $q_i=|C_i|$. Let $g$ and $h$ be the quotient and remainder of  $\frac{D}{\sum_{i=1}^{l}{T_i}}$ respectively, and $t_0 + h$ is the time where the network topology is $G_p$ for a $p$, $1\le p\le l$. Therefore, the topologies in time duration $t_0$ to $t_0+D$ includes $G_1$ to $G_l$ for $g$ number of cycles and $G_1$ to $G_p$ in last periodic cycle. Starting from first topology $G_1$ in first period, in each topology $G_i$, if a node $v$ is in the same connected component with a node $w\in N(u)$, then $v$ can be reachable from $u$ through a temporal path which is completed in $G_i$; hence, $N(u)$ is updated to $N(u)\cup (\bigcup_{k: N(u)\cap C_{ik}\neq \emptyset} C_{ik})$. In this step the algorithm goes through all the topologies from $t_0$ to $t_0+D$. In the end, if $N(u)= V$ for all $u\in V$ then the transmission range $Tr$ is sufficient for having a connected network with delay $D$. In Algorithm \ref{Alg:DTNconnectivity} the steps of checking the connectivity of a dynamic graph with delay $D$ is proposed.

\begin{algorithm}[H]
\caption{Checking Connectivity of Airborne Network with delay $D$}
\label{Alg:DTNconnectivity}
{\em Input}: $\mathcal{G}(t)=\{G_1, G_2, \ldots, G_l\}$ and delay threshold $D$\\
{\em Output}: {\em true} if dynamic graph $\mathcal{G}(t)$ is connected with delay $D$; otherwise {\em false}.


\begin{algorithmic}[1]
\STATE Initialize $N(u) = \{u\}$ for every $u\in V$
\STATE \textbf{for all} {topologies $G_i, 1\le i \le l$}
\STATE  \ \ \ Compute $C_i=\{C_{i,1}, C_{i,2},\ldots C_{i,q_{i}}\}$, the set of connected components of $G_i$
\STATE \textbf{for all} {periods 1 to $g$}
\STATE \ \ \  \textbf{for all} {topologies $G_i, 1\le i \le l$}
\STATE \ \ \ \ \ \ \textbf{for all} {node $u\in V$}
\STATE \ \ \ \ \ \ \ \ \ $N(u)\leftarrow N(u)\cup (\bigcup_{k: N(u)\cap C_{i,k}\neq \emptyset} C_{i,k})$
\STATE \textbf{for all} {topologies $G_i, 1\le i \le p$ (the topologies in the last period)} 
\STATE \ \ \  \textbf{for all} {node $u\in V$}
\STATE \ \ \ \ \ \ $\displaystyle N(u)\leftarrow N(u)\cup (\bigcup_{k: N(u)\cap C_{i,	k}\neq \emptyset} C_{i,k})$

\STATE \textbf{for all} {node $u\in V$}
\STATE \ \ \ {\bf if} $N(u)\neq V$, {\bf return} $false$
\STATE {\bf return} $true$
\end{algorithmic}	
\end{algorithm}    

 \vspace{-0.2 in}
As we explained in section \ref{sec:dynamictopology}, number of topologies, $l$ in one period is $O(n^2)$. The computation of the connected components of a graph $G_i=(V,E_i)$ needs using either breadth-first search or depth-first search with time complexity of $O(|V| + |E_i|)=O(n^2)$. Hence, step 2-4 takes $O(n^4)$. This algorithm is used for the case that $D < (l-1)\sum_{i=1}^{l}{T_i}$. Therefore, number of periods $g < l-1$ and $g=O(n^2)$. Computation of Step 7 also needs $O(n^2)$ since $|N(u)|$ and total size of all components in $G_i$ is $O(n)$. Finally, we can conclude that total time complexity of the algorithm is $O(n^7)$.
 
As we mentioned before, in Algorithm \ref{Alg:DTNconnectivity} the delays are computed with respect to $t_0$. We can easily extend it to any time in the network operation duration, by repeating Algorithm \ref{Alg:DTNconnectivity} for every $t_i, 1\le i\le l$ where $t_i$ is the starting time of topology $G_i$. The complexity increases by a factor of $l=O(n^2)$. We note that in this case even if a node starts communication at some time instances $t$ where $t_i\le t\le t_{i+1}$, the delay will be smaller than the case it starts at $t_i$. Hence it is enough to just consider the time points in which a topology change happens. 

Similar to the computation of CTR and CTR$_f$, in order to compute CTR$_D$ we can conduct a binary search within the range $0 - Tr_{max}$ and we can determine the smallest transmission range that will ensure the AN is connected with delay $D$ during the entire operational time. The binary search adds a factor of $\log Tr_{max}$ to the complexity of Algorithm \ref{Alg:DTNconnectivity}.

\section{Simulations}
\label{sec:simulation}

The goal of our simulation is to compare critical transmission range in different scenarios of non faulty, faulty and delay tolerant and investigate the impact of various parameters, such as the number of ANPs, the region radius and delay on critical transmission range. 
 In our simulation environment, the deployment area is a 1000 $\times$ 1000 square mile area. The centers of the orbits of the ANPs are chosen randomly in such a way that the orbits do not intersect with each other. In our simulation, we assume that all the ANPs move at the same angular speed of $\omega$ = 20  radian/hour. Hence a period length is $0.1\pi$ hour. One interesting point to note is that, in this environment where all the ANPs are moving at the same angular speed on circular paths, the value of $CTR$ is independent of the speed of movement of the ANPs. This is true because changing the angular speed $\omega$ effects just the time at which the events, such as a link becomes active or a link dies, take place. If we view the dynamic topology of the backbone network over one time period as a collection of topologies ${\cal G} = \{G_1, G_2,  \ldots, G_l\}$, where $G_i$ morphs into $G_{i+1}, 1 \leq i \leq l$ at some time, by increasing or decreasing the angular speed of all the ANPs, we just make the transitions from $G_i$ to $G_{i+1}$ faster or slower, without changing the topology set $\cal G$. Similarly, the set of ANPs that are damaged to failure of a region at a certain time, remains unchanged.

\begin{figure}[htb]
\vspace{-0.1in}
 \centering
\includegraphics[width=0.5\textwidth,keepaspectratio]{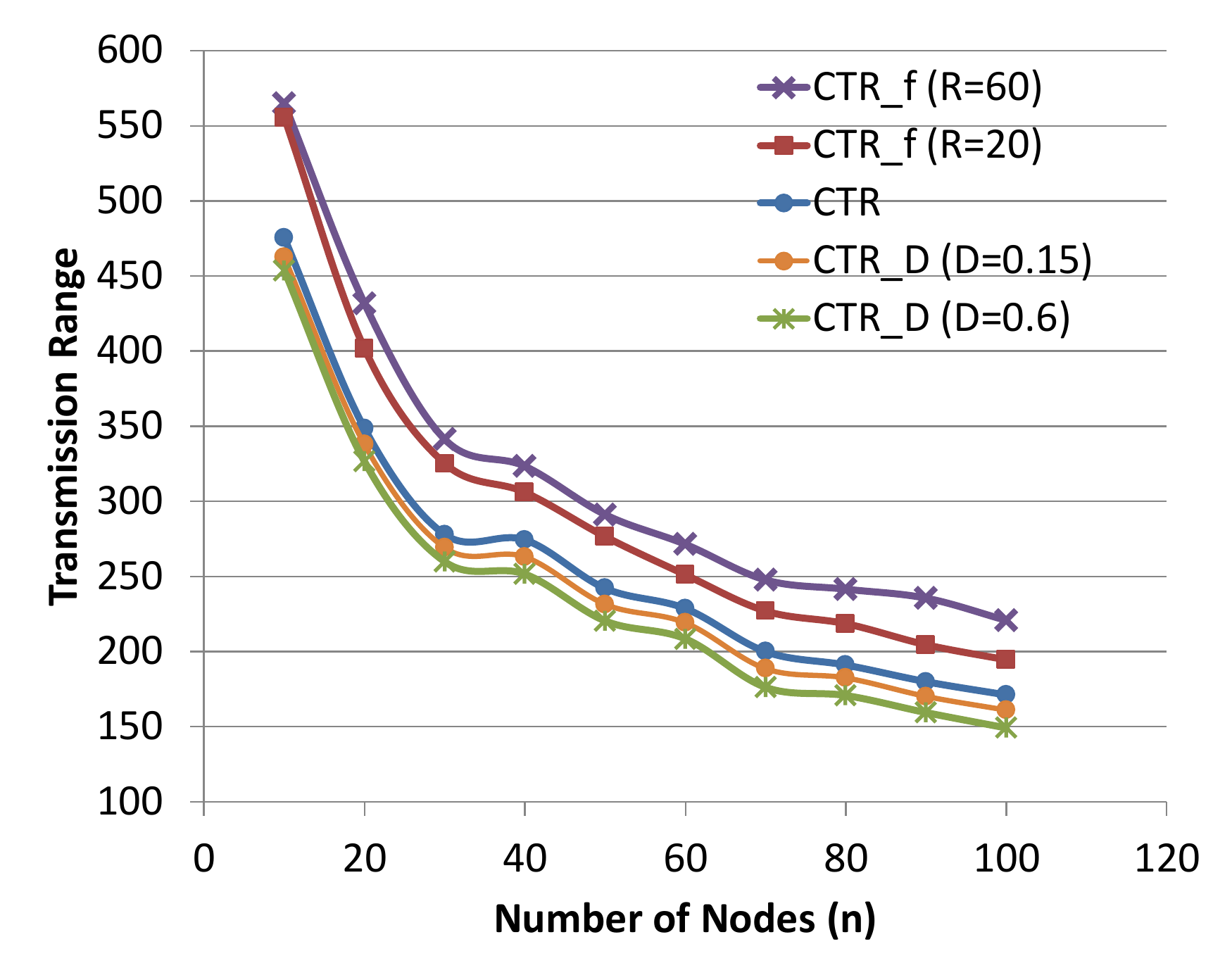}
\vspace{-0.2in}
\caption{Transmission Range vs. Number of Nodes}
\vspace{-0.2in}
\label{TrN}
\end{figure}

In our first set of experiments we compute $CTR$, $CTR_f$ when $R=20,60$, and $CTR_D$ when $D=0.5period, 2period$ for different values of number of nodes, $n$. Fig. \ref{TrN} depicts the result of these experiments. In these experiments, for each value of $n$ we conducted 30 experiments and the results are averaged over the 30 different random initial setups. We set $orbit\ radius=10$. We observe that expectedly an increase in the number of nodes results in a decrease in $CTR$, $CTR_f$ and $CTR_D$. Moreover, $CTR_D\le CTR\le CTR_f$ for all instances. In all of the experiments, we compute $CTR_D$ with respect to all times (corresponding to beginning of a new topology) not only $t_0$.   

\begin{figure}[htb]
 \centering
  \centering
 \subfigure[]{
\includegraphics[width=0.4\textwidth,keepaspectratio]{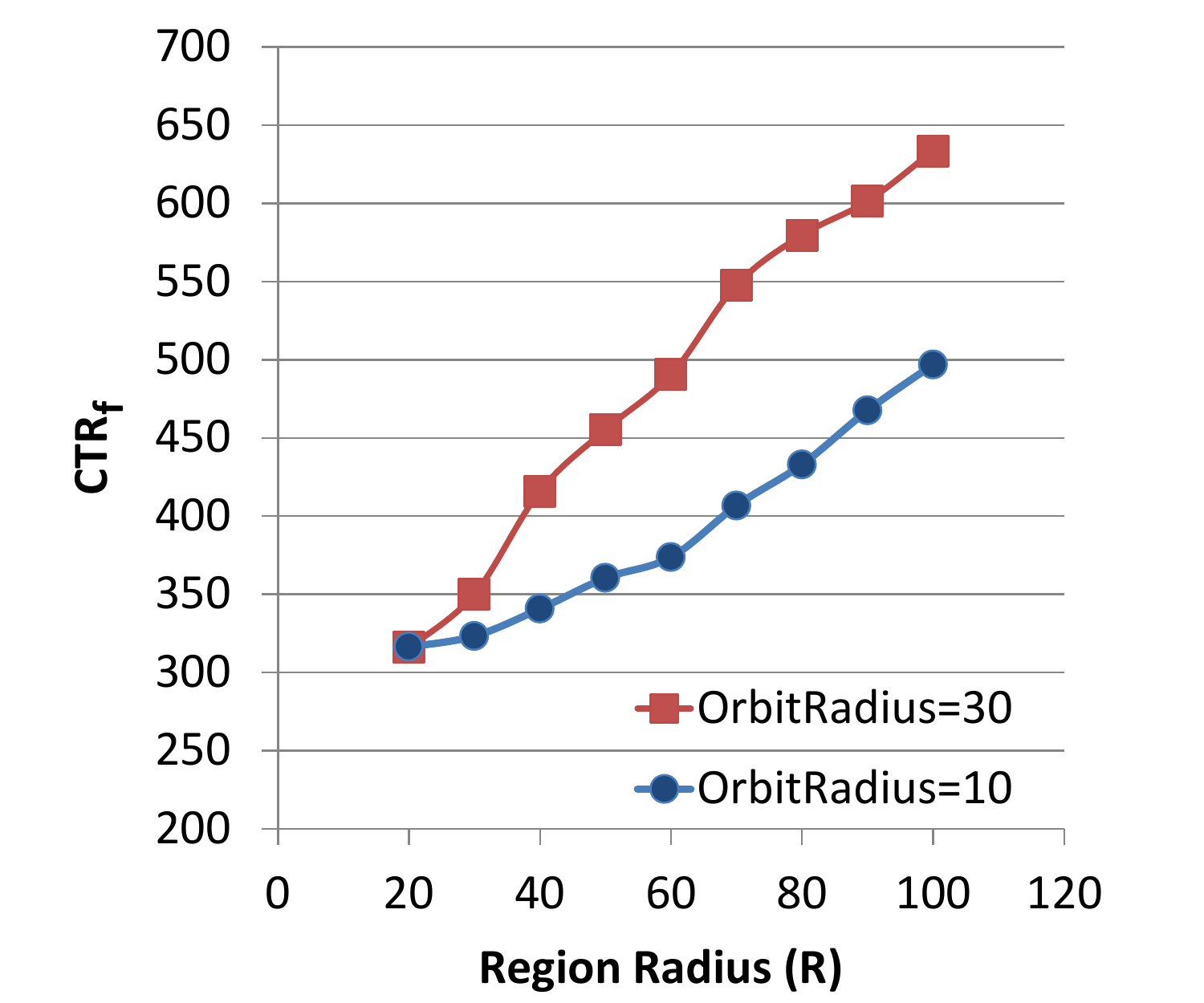}
\label{TrR}}
 \subfigure[]{
 \includegraphics[width=0.4\textwidth,keepaspectratio]{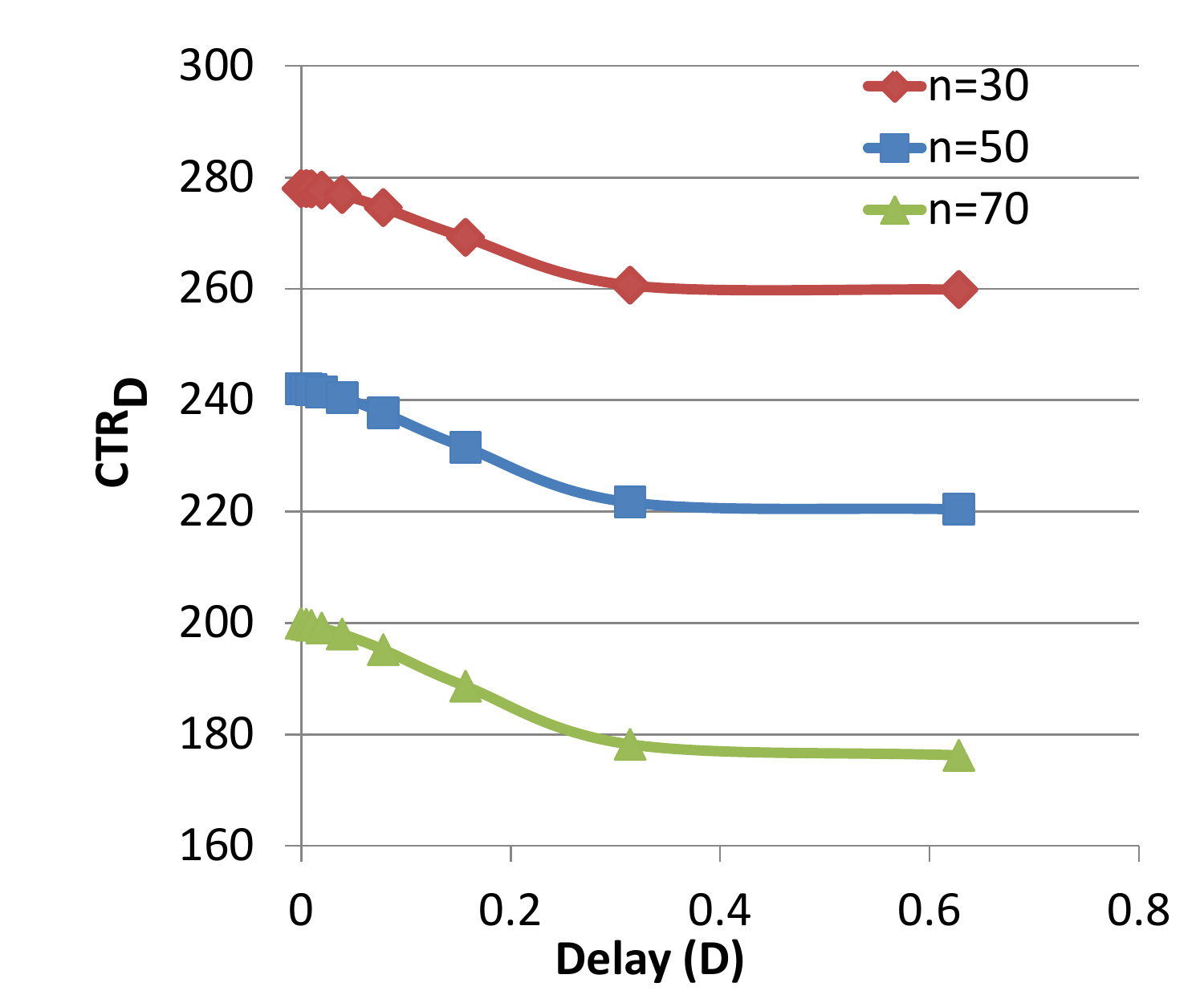}
 \label{delayPlot}}
\caption{(a) Transmission Range ($CTR_f$) vs. Region Radius, $n=35$; (b) Transmission Range ($CTR_D$) vs. Delay}
\label{CTRDandCTRf}
\end{figure}

In the second set of experiments, we examined the impact of change of the region radius $R$ on the transmission range. We conducted these experiments for two values of $orbit\ radii$, 10 and 30, and $n = 35$ in both the cases. For each value of $R$, we conducted 100 experiments and the results are averaged over them. Fig. \ref{TrR} shows the results. It may be observed that increase in the value of $R$ leads to increase in CTR. 
 This observation is quite expected as larger regions can destroy more nodes at a time. Moreover, it may be noted that for larger values of orbit radii the transmission range also increases. The reason is that for a specific number of nodes in a bounded deployment area, larger orbit radii result in larger distance between the nodes. Accordingly, larger transmission range is necessary,  particularly in the case of larger $R$s.

Finally, we conducted experiments to investigate the impact of delay $D$ on the value of $CTR_D$. Fig. \ref{delayPlot} depicts the results. We observe that when value of delay $D$ is zero the value of $CTR_D$ is equal to $CTR$ and by increasing delay, $CTR_D$ decreases and the interesting observation is that when delay becomes greater than $2 period$ the decrease in the value of $CTR_D$ is unnoticeable or even zero.   

\section{Conclusion}
\label{sec:conclusion}

Existence of sufficient control over the movement pattern of the mobile platforms in Airborne Networks opens the avenue for designing topologically stable airborne networks. In this paper, we discussed the system model and architecture for Airborne Networks (AN). We studied the problem of maintaining the connectivity in the underlying dynamic graphs of airborne networks when trajectories of nodes are given. We developed techniques to compute the {\em dynamic topology} of the AN at any instance of time and proposed an algorithm to compute critical transmission range when all nodes are operational. 
 Motivated by the importance of robustness and fault tolerance capability of ANs, we have also investigated the region-based connectivity of the ANs and proposed an algorithm to find the minimum transmission range necessary to ensure that the surviving nodes of the network remain connected, even when all or some nodes of region fail due to an enemy attack. In the process of computing the minimum transmission range in faulty scenario, we developed techniques to (i) compute all the fault regions that need to be considered to {\em ensure overall connectivity at all times} and (ii) compute the set of  nodes that might be damaged by the {\em failure of a specific region at a specific time}. 
 Moreover, we defined and formulated the critical transmission range in delay tolerant airborne networks $CTR_D$ and proposed an algorithm to compute $CTR_D$.   
Through simulations, we have illustrated the impact of the number of nodes, the region radius in faulty scenario and delay in delay tolerant networks on the minimum transmission range. In future we plan to develop more efficient algorithms to compute critical transmission range in different scenarios and also to study the environment where the ANPs take unpredictable flight paths. 

\bibliographystyle{IEEEtran}
\bibliography{IEEEabrv,references}

\end{document}